\documentclass[reprint,superscriptadress,showpacs,preprintnumbers,amsmath,amssymb,aps,prl]{revtex4-1}
\usepackage[utf8]{inputenc}

\usepackage{graphicx}

\begin{document}

\title{Ultrafast dynamics of photocurrents in surface states of 3D topological insulators}

\author{Jens G{\"u}dde}
\affiliation{
Fachbereich Physik und Zentrum f\"{u}r Materialwissenschaften,
Philipps-Universit\"{a}t Marburg, D-35032 Marburg, Germany
}
\author{Ulrich H{\"o}fer}
\affiliation{
Fachbereich Physik und Zentrum f\"{u}r Materialwissenschaften,
Philipps-Universit\"{a}t Marburg, D-35032 Marburg, Germany
}

\begin{abstract}

This article reviews experimental work on the ultrafast electron
dynamics in the topological surface state (TSS) of three-dimensional
(3D) topological insulators (TIs) observed with time- and
angle-resolved two-photon photoemission (2PPE).
 The focus is laid on the generation of ultrafast photocurrents and the
time-resolved observation of their decay.
 2PPE not only allow to unambiguously relate the photocurrents to the
spin-polarized electronic surface states.  Probing of the asymmetric
momentum distribution of the electrons carrying the current makes it
possible to study the microscopic scattering processes that governs
the unusual electron transport in the time domain.
 Ultrashort mid-infrared pump pulses permit not only a direct optical
excitation of the TSS in Sb$_2$Te$_3$ but also lead to a strong
asymmetry of the TSS population in momentum space.
 Two-dimensional band mapping of the TSS shows that this asymmetry is
in fact representative for a macroscopic photocurrent while the
helicity-dependence of the photocurrent is found to be small.
 The time-resolved observation of the photocurrent decay reveals a huge
mean free path of the electrons in the TSS.

\end{abstract}

\maketitle


\section{Introduction}

\subsection{Basic properties of three-dimensional topological insulators}
Three-dimensional topological insulators have attracted a
lot of attention because they are characterized by an insulating bulk
and a conducting surface state with remarkable properties which are
interrelated and induced by strong spin-orbit coupling in combination
with time-reversal symmetry \cite{Fu07prl,Xia09natphys,Zhang09natphys,Hasan10rmp}.
 One of these properties is its topological protection which guaranties
the existence of the topological surface state as long as
time-reversal symmetry is not broken.
 This makes the TSS robust against nonmagnetic pertubations
which is beneficial for application in real devices.
 Moreover, the TSS promises unique electron transport properties due
its linear Dirac-like energy-momentum dispersion in combination with spin-momentum
locking which results in a chiral spin structure in momentum space.
 On the one hand, this implies that electrical currents in the TSS are
automatically spin-polarized because opposite linear momenta are
linked to opposite spin orientations \cite{Hsieh09nat}.
 On the other hand, the linear dispersion results in a small phase
space for inelastic electron-electron scattering of excited electrons
as in graphene \cite{Gonzales96prl} while the chiral spin structure
additionally suppresses momentum scattering with complete absence of
direct backscattering \cite{Roushan09nat}.
 This means that spin-polarized electric currents in the TSS are
expected to flow ballistically over large distances which makes these
surface states promising for use in ultrafast low-loss electronics and
spintronics.

The verification and application of these unique transport properties,
however, turned out to be rather difficult because this requires that
the transport is in fact dominated by the TSS. This needs TIs with a
sufficiently large bulk band gap and a TSS that covers this gap.
 Beside the first predicted 3D TI Bi$_{1-x}$Sb$_x$ \cite{Fu07prb},
tetradymites such as the binary chalcogenides Bi$_2$Se$_3$, Bi$_2$Te$_3$, and
Sb$_2$Te$_3$ belong to the same class of three-dimensional
topological insulators and are characterized by a single Dirac cone
around $\bar{\Gamma}$ and a relatively large indirect band gap \cite{Xia09natphys,Zhang09natphys}
compared to other known 3D topological insulators.
 The largest band gap is found in Bi$_2$Se$_3$ but amounts to only about 300~meV.
 This makes these materials in fact to narrow-gap semiconductors and
their transport properties depend sensitively on the actual position
of the Fermi level E$_F$ which is often governed by unintentionally doping.
 For this reason, transport measurements on TIs are commonly dominated by their
bulk rather than their unusual surface properties.
 Bi$_2$Se$_3$, for example, is found to be highly n-doped due to Se vacancies
\cite{Navratil04jssc,Hor09prb} and its transport properties are dominated
by the bulk conduction band.
 On the other hand, the corresponding shift of the energy of the Dirac point
$E_{\rm D}$ below the Fermi energy $E_{\rm F}$ makes the topological
surface state of Bi$_2$Se$_3$ accessible by angle-resolved
photoelectron spectroscopy (ARPES).
 This made ARPES to the most important tool for the characterization of
TIs and made it not only possible to verify the linear dispersion of
the TSS very shortly after its prediction
\cite{Xia09natphys,Chen09sci2,Hsieh08nat} but also to proof its
chiral spin texture by combination of ARPES with spin-resolved
detection \cite{Hsieh09nat}.

\subsection{Mechanisms for photocurrent generation}
The generation of a photocurrent, in particular by using ultrashort
laser pulses, opens the possibility to investigate the unusual transport
properties of TIs in the time domain and on the time scale of the scattering
processes that governs the charge transport.
 Several mechanisms and techniques for the optical generation of dc
electrical currents in condensed matter are known which are
particularly well established for the generation of photocurrents in
the conduction band of bulk direct-band-gap semiconductors.
 This includes a scheme that additionally allows for a coherent control
of the size and direction of the photocurrent by tuning the relative
phase of two phase-locked laser fields with photon energies
$\hbar\omega$ and $\hbar\omega/2$, which simultaneously drive a one-
and two-photon transition, respectively.
 This method has been at first used to induce electric
\cite{Dupont95prl,Atanasov96prl,Hache97prl} and spin currents
\cite{Stevens03prl,Hubner03prl} in bulk direct-band-gap semiconductors
and has been combined with time- and angle-resolved 2PPE to
investigate the ultrafast dynamics of electric currents in
image-potential states at a Cu(001) surface~\cite{Gudde07sci}.
 Recently, it has been demonstrated for epitaxially thin films
  of Bi$_2$Se$_3$ using photon energies that exceed the bulk band gap
  \cite{Bas15apl,Bas16oe}.
The generation of a dc current by this process can be understood as
third-order nonlinear optical rectification which allows to generate
photocurrents even in unbiased centrosymmetric
materials~\cite{Aversa95prb}.
 An exclusive current generation in a TSS by this
method~\cite{Muniz14prb}, however, requires phase-stable two-color
coherent control at very low photon energies because of the small band
gap of typical topological insulators.

 Other methods for photocurrent generation are based on single-color
excitation schemes such as the photogalvanic effect
~\cite{Ganichev03jp,Ivchenko05,Ganichev06}.
 The PGE is a second-order nonlinear effect and is therefore forbidden
in inversion symmetric systems such as the bulk of the tetradymites
B$_2$Se$_3$, Bi$_2$Te$_3$ and Sb$_2$Te$_3$ which share the same
rhombohedral crystal structure with the space group $D_{3d}^5$
\cite{Zhang09nat}.
 This makes it possible to generate photogalvanic currents in these
materials by a single-color excitation exclusively at the surface
where the inversion symmetry is broken.
 The helical spin structure of the TSS of three-dimensional TIs
motivates to utilize the circular photogalvanic effect to
generate a directional helicity-dependent photocurrent because the
coupling of circular polarized light depends on the spin orientation
of the surface electrons with respect to the photon momentum.
 For surfaces with rotation symmetry, such as the threefold symmetry of
the (111) surface of the tetradymite TIs, the circular photogalvanic
effect vanishes at normal incidence even if the spin orientation has
an out-of-plane component~\cite{Hosur11prb} because the latter results
in a threefold symmetric excitation which does not generate a net
directional current \cite{Huang20sr}.
 At oblique incidence, however, a helicity-dependent photogalvanic
current can be generated perpendicular to the plane of
incidence~\cite{McIver12natnano,Duan14scirep,Huang20sr}.
 A photogalvanic current can be also generated by linear polarized light,
which is described by the linear photogalvanic effect.
 At the threefold symmetric (111) surface of the tetradymite TIs, the
direction of a photocurrent generated by the linear photogalvanic
effect depends on the orientation of the light polarization with
respect to the mirror plane of the surface.
 In contrast to the circular photogalvanic effect, it is expected 
that the linear photogalvanic effect has its maximum magnitude at normal
incidence \cite{Olbrich14prl}.
 For small frequencies that cannot induce interband transitions, the
linear photogalvanic effect is microscopically described by asymmetric
scattering of the excited electrons that are accelerated by the
oscillating electric field \cite{Olbrich14prl}.

Junck {\it et al.} have theoretically studied the possibility to
generate photocurrents through the photogalvanic effect by optical
transitions between the occupied and unoccupied part of an isolated
TSS and found that pure orbital coupling can neither generate a
helicity-dependent nor a helicity-independent photocurrent
\cite{Junck13prb}.
 Only if they include the very small Zeeman coupling between the spins
of the electrons and the magnetic field of the light, they can predict
a helicity-dependent photocurrent that is, however, very small because
it scales with the square of the Zeeman coupling.

 A competing process that is even allowed in the bulk of centrosymmetric
media is the photon drag effect which describes photocurrent
generation by momentum transfer from light to carriers
\cite{Grinberg70jetp,Ivchenko05,Ganichev06}.
 It is not straightforward to distinguish the photogalvanic effect and
the photon drag effect in experiments, because both processes share
the same dependence on the light polarization.
 In contrast to the photogalvanic effect, however, a photocurrent
induced by the photon drag effect changes its sign when the direction
of light propagation is reversed.
 An unambiguous identification of surface photocurrents in the
tetradymite TIs without a surface specific probe therefore requires
the comparison of data taken under front and back side illumination
\cite{Olbrich14prl,Plank16prb}.

\subsection{Two-photon photoemission (2PPE)}
In contrast to conventional ARPES, which is restricted to the
spectroscopy of occupied electronic states, two-photon photoemission
(2PPE) can additionally access initially unoccupied states by
populating them first with a short excitation pulse and photoemitting
the excited electrons with a second pulse
\cite{Haight95ssr,Fauster95}.
 This makes it not only possible to investigate the unoccupied TSS of
intrinsically p-doped samples. By introducing a variable time-delay
between these two pulses, 2PPE can be combined with ultrafast
pump-probe spectroscopy \cite{Haight95ssr,Bovens10}.
 It has been also early combined with angle-resolved detection
\cite{Haight95ssr,Berthold02prl,Rohleder05njp} and is then nowadays often
called time-resolved ARPES (trARPES).
 This term, however, might suggest that the photoemission step only
subsequently images the transiently excited population while the 2PPE
process also includes coherent effects of the two-photon transition
\cite{Petek97pss,Gudde06apa,Reutzel20natcomm}.

\begin{figure}
 \includegraphics[width=\columnwidth]{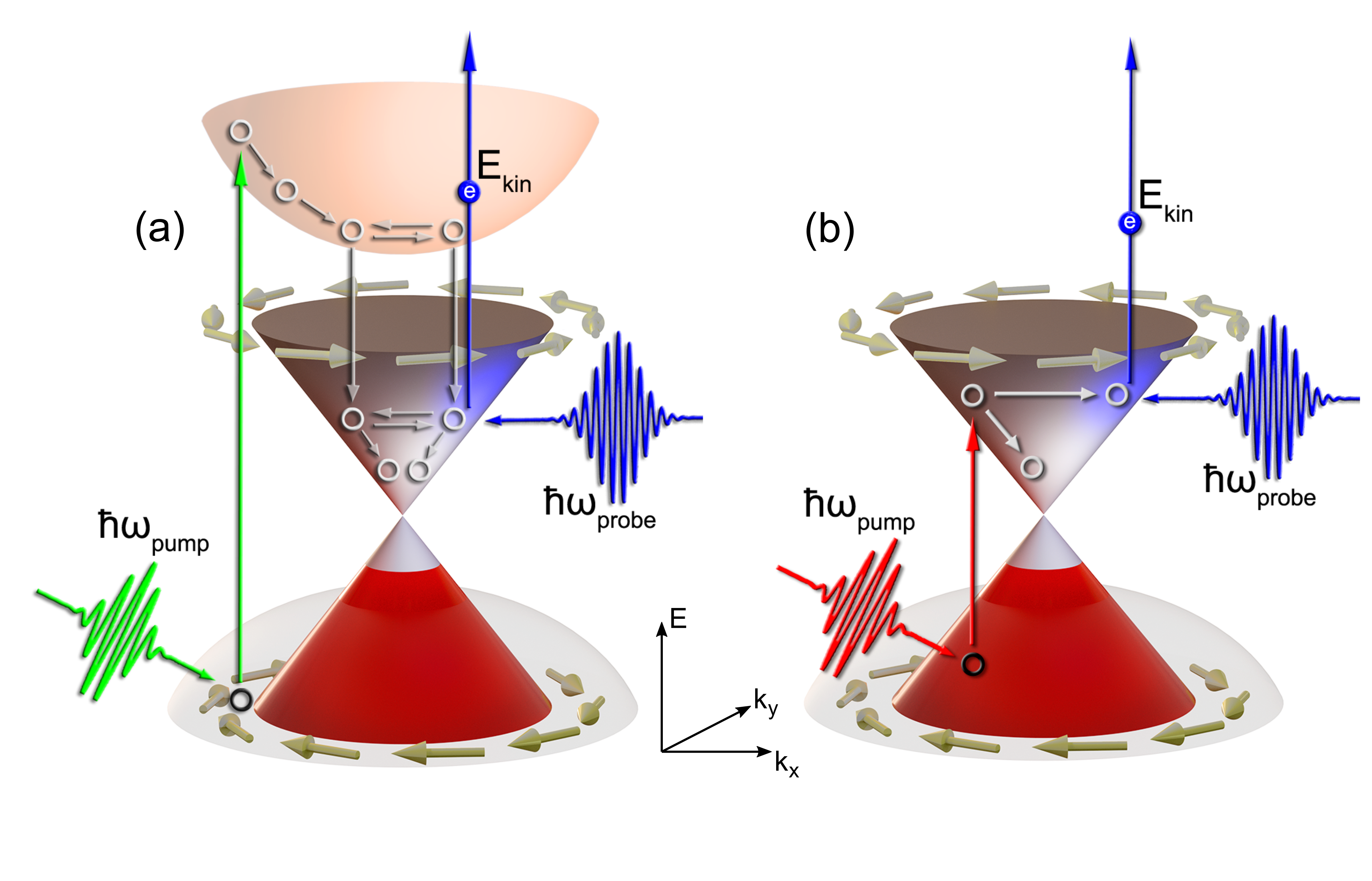}
 \caption{Scheme for 2PPE of a partial filled TSS which is characterized
   by a Dirac-cone-like energy dispersion in the two-dimensional momentum
   space ($k_x$,$k_y$) of the surface and a chiral spin structure
   as indicated by the circulating arrows.  (a)
   Pump pulses with photon energies that exceed the bulk band gap
   initially excite electrons in electronic bands far above the TSS.
   This results in a delayed filling of the TSS under participation of
   many scattering events that homogenize the momentum distribution
   in the TSS even if the initial excitation might have induced an
   asymmetry in momentum space, i.e. a photocurrent. (b) Direct optical
   excitation of the unoccupied part of the TSS with mid-IR pump
   pulses. The subsequent photoemission images the initially generated
   momentum distribution and monitors its redistribution and decay
   on an ultrafast time scale.}
  \label{fig:2ppe-scheme}
\end{figure}

Time- and angle-resolved 2PPE is even capable to investigate electron
transport on an ultrafast time scale if the initial optical excitation
by the pump pulses creates an asymmetry in the momentum distribution
of the electrons parallel to the surface, i.e. a photocurrent
\cite{Gudde07sci}.  The time-resolved observation of the
redistribution and decay of the initially inhomogeneous momentum
distribution provides microscopic information on the different
scattering mechanisms of just those electrons that carry the
photocurrent.

 In the following, we will first review 2PPE experiments of TIs that
utilize visible or near-infrared light for excitation.
 In section~\ref{sec:mid-IR-excitation}, we will then show that
excitation of Sb$_2$Te$_3$ by linear polarized mid-infrared (mid-IR)
pulses generates an inhomogeneous population of the TSS in momentum
space (section~\ref{subsec:direct-excitation}) that can be
unambiguously identified as a persistent macroscopic photocurrent by
angle-resolved 2PPE covering both momentum directions of the surface
band structure (section~\ref{subsec:photocurrent-generation}).
 This photocurrent is most pronounced if the excitation breaks the
threefold symmetry of the Sb$_2$Te$_3$(0001) surface, i.e. if the
plane of oblique light incidence is aligned perpendicular to a mirror
plane of the surface and vanishes for the opposite orientation.
 In spite of the helical spin texture of the TSS, the helicity
dependence of the photocurrent is found to be small but can control
the magnitude and sign of the photocurrent for the latter sample
orientation (section~\ref{subsec:polarization-control}).
 Decomposing the decay dynamics of the photocurrent into inelastic
electron scattering to lower energies and elastic momentum scattering
within the TSS reveals that the ballistic mean free path of the Dirac
fermions reaches almost 1~$\mu$m resulting from the suppression of
backscattering in the TSS (section~\ref{subsec:decay-dynamics}).
 A similar large mean free path is found in Bi$_2$Te$_3$ by a novel
combination of THz acceleration of the Dirac fermions and
time-resolved ARPES, which makes it possible to investigate electron
transport just at the Fermi level with subcycle time resolution as is
reviewed in section~\ref{sec:thz}.
 These results present an unambiguous experimental verification of the
unusual transport properties of the TSS of 3D topological insulators
resulting form spin-momentum locking.

\begin{figure}
 \includegraphics[width=\columnwidth]{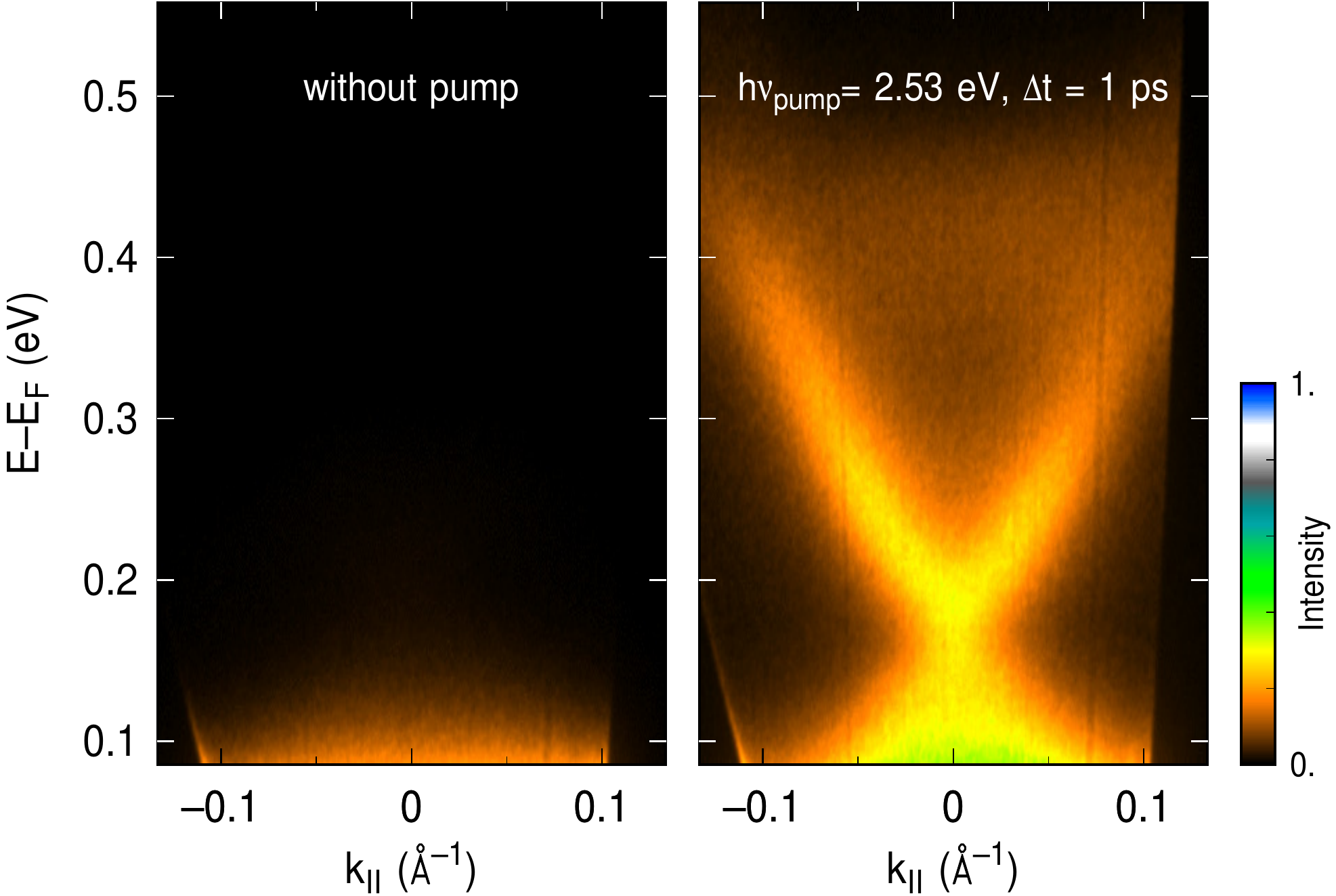}
 \caption{Angle-resolved 2PPE spectra of Sb$_2$Te$_3$. The electrons
   were collected along the $\bar{\Gamma}$-$\bar{K}$ direction,
   perpendicular to the plane of light incidence. A probe
   photon energy of 5.06~eV suppresses photoemission of the initially
   occupied states below $E_F$ and enhances the dynamic range of the
   2PPE signal. (a) without pump pulses.  (b) 1~ps after arrival of
   the 2.53-eV pump pulses. }
  \label{fig:2ppe-sb2te3}
\end{figure}

\section{Visible and near-IR excitation}
\label{sec:visible-excitation}

Most 2PPE studies on the TSS of TIs used pump pulses in the
near-infrared or visible region
\cite{Wang12prl,Crepaldi12prb,Hajlaoui12nl,Sobota12prl,Niesner12prb,
  Crepaldi13prb,Hajlaoui13epj,Sobota13prl,Niesner14prb,Hajlaoui14natcomm,Reimann14prb,
  Neupane15prl,Zhu15sr2,Sanchez16prb,Jozwiak16natcomm,Sumida17scirep,
  Bugini17jpcm,Sanchez17prb,Ketterl18prb,Soifer19prl}.
Their photon energy therefore substantially exceeds the bulk band gap
of the TIs. This typically prohibits a direct optical excitation of
the TSS because the pump pulses predominantly excite electronic states
with energies far above the TSS.
 Depending on the excitation conditions, this can include transitions
for all combinations of bulk and surface initial and final states.
 These higher lying states decay subsequently to lower lying states which results in
a delayed filling of the TSS \cite{Sobota12prl,Reimann14prb} as is
depicted in Fig.~\ref{fig:2ppe-scheme}(a).  This typically prevents
the generation of an asymmetry in the momentum distribution of the TSS
and therewith a spin polarized photocurrent. Even if the initial
excitation might induce such asymmetry in higher lying states
\cite{Soifer19prl}, the sequential decay into the TSS close to the
Fermi level goes along with multiple scattering events that
homogenizes the momentum distribution.

\begin{figure*}
 \includegraphics[width=0.85\textwidth]{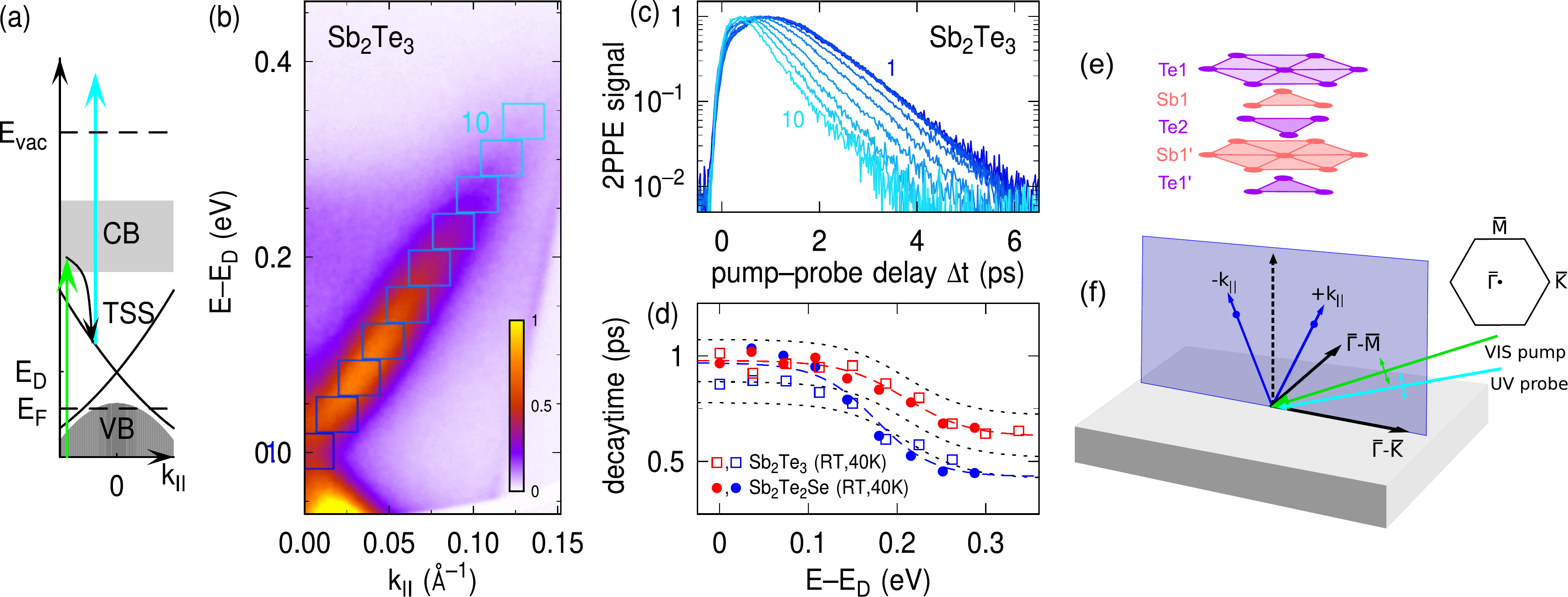}
  \caption{(a) Excitation scheme for the population of the TSS with
    visible pump pulses via transfer from the bulk conduction band and
    subsequent photoemission with ultraviolet probe pulses. (b) 2PPE
    spectrum of the TSS of Sb$_2$Te$_3$ along the along the
    $\bar{\Gamma}$-$\bar{K}$ direction. The rectangles indicate the
    integration windows for the transient 2PPE intensities. (c)
    Transient 2PPE intensities of Sb$_2$Te$_3$ at room temperature
    (RT) within the integration windows shown in (b). (d) Decay times
    of the transient 2PPE intensities for Sb$_2$Te$_3$ (open squares)
    and Sb$_2$Te$_2$Se (filled dots) at RT (red) and 40 K (blue),
    respectively. The dashed lines are guides to the eye. The black
    dotted lines indicate the variation of the decay times for
    different cleaves. Adapted with permission.\cite{Reimann14prb}
    Copyright 2014, American Physical Society.  (e) Structure of a
    quintuple layer of Sb$_2$Te$_3$ according to
    Ref.~\cite{Zhang09natphys}.  (f) Experimental geometry with sample
    orientation and surface Brillouin zone. In these experiments, the
    electrons were collected along the plane of light incidence.}
  \label{fig:reimann14prb}
\end{figure*}

While first 2PPE studies on TIs were investigating Bi$_2$Se$_3$
\cite{Wang12prl,Crepaldi12prb,Sobota12prl,Crepaldi13prb,Hajlaoui13epj,Sobota13prl}
or Bi$_2$Te$_3$ \cite{Hajlaoui12nl,Hajlaoui14natcomm}, the work presented here is
focused on the third prototype 3D TI Sb$_2$Te$_3$.
 Sb$_2$Te$_3$ is intrinsically p-doped, which is attributed to
substitutional Sb defects at Te sites \cite{Horak88pss}, and most of
the TSS is unoccupied.
 This makes it difficult to investigate the TSS by conventional ARPES
\cite{Pauly12prb,Seibel12prb} and the Dirac-cone-like electronic
structure of this prototype TI could until then only be verified
indirectly by Landau level spectroscopy \cite{Jiang12prl}.
 However, it makes Sb$_2$Te$_3$ an ideal system to study the electron
dynamics within a TSS by 2PPE.
 The first 2PPE experiments on Sb$_2$Te$_3$ of Reimann {\it et al.}
\cite{Reimann14prb} have for the first time directly revealed that the
massless Dirac-cone like energy dispersion of TSSs is in fact also
realized in Sb$_2$Te$_3$ above the Fermi level as shown in
Fig.~\ref{fig:2ppe-sb2te3}.

 This work furthermore compared the electronic structure and electron
dynamics of Sb$_2$Te$_3$ with Sb$_2$Te$_2$Se because theoretical
predictions have suggested that the substitution of the central Te
layer in the quintuple layers of Sb$_2$Te$_3$ (Te2 in Fig.~\ref{fig:reimann14prb}(e))
by Se or S preserves the
topological surface state, but considerably increases the bulk band
gap \cite{Menshc11jetpl,Lin11njp}.
 This should result in a better decoupling of the TSS from the bulk
bands and therefore improve the transport properties.
 The 2PPE data on the dispersion of the TSS in both Sb$_2$Te$_3$ and
Sb$_2$Te$_2$Se are in very good agreement with the band structure
calculations and in fact indicate that the parabolic bulk conduction
band (BCB) of Sb$_2$Te$_2$Se is shifted towards higher energies as
compared to Sb$_2$Te$_3$.

The electron dynamics in Sb$_2$Te$_3$ and Sb$_2$Te$_2$Se at room
temperature and at 40~K was compared in order to identify the main
decay mechanisms of electrons in the TSS.
 Fig.~\ref{fig:reimann14prb}(c) exemplarily shows transient data for
Sb$_2$Te$_3$ employing visible (2.58 eV) pump light.
 The slow rise of the transients reflects the delayed
filling from higher lying states and the maximum population
of the TSS is reached only up to 1 ps after the optical
excitation.
 The decay of the 2PPE signal for larger delays can be well described
by a single exponential.
 Two decay channels have been identified which are both related to the
coupling of the TSS with bulk states.
 The dominating one is electron-hole pair creation in the partially
filled valence band as has been also concluded for other p-doped
materials \cite{Hajlaoui14natcomm,Niesner14prb}.
 The decay time of electrons in the TSS therefore depends strongly on
the number of unoccupied states in the valence band and it has been
shown that it can be enhanced by up to two orders of magnitude by
tuning the Fermi level through doping of Sb$_2$Te$_3$ with Bi
\cite{Sumida17scirep}.
 In bulk insulating Bi$_2$Te$_2$Se, a decay time of even more than
4~$\mu$s has been observed \cite{Neupane15prl}.
 The second important decay channel is electron transport out of the
probed volume close to the surface into the bulk through coupling of
the TSS to the bulk conduction band.
 As can be seen in Fig.~\ref{fig:reimann14prb}(d), the decay time
gets shorter at higher energies just where the TSS starts to degenerate
with the bulk conduction band.
 This reduction of the decay time at higher energies is even stronger at
40~K which can be explained by the higher electron mobility in bulk
Sb$_2$Te$_3$ at lower temperatures.

 While these experiments mainly probed the electron transport into the
bulk, the investigation of the unusual transport properties within the
TSS by 2PPE requires at first the optical generation of an inhomogeneous
momentum distribution in the TSS whereat an asymmetry along a certain
direction corresponds to a photocurrent.

\subsection{Helicity-dependent photocurrents on Bi$_2$Se$_3$}
 The first experiment that found a helicitiy-dependent photocurrent on
a 3D TI was performed by McIver {\it et al.}  \cite{McIver12natnano}
on Bi$_2$Se$_3$ using laser light with a photon energy of 1.56~eV
\cite{McIver12natnano}.
 The laser was focussed onto the sample in the center between two
metallic contacts that picked up the induced electrical current and
the helicity of the light was continuously varied by a
$\lambda/4$-plate.
 Beside a helicity-independent thermoelectric current due to
inhomogeneous laser heating, a helicity-dependent photocurrent was
found, the dependence on light helicity and angle of incidence of
which just fitted the behavior expected for the circular photogalvanic effect.
 The photon drag effect has been ruled out because of the bulk spin degeneracy
of Bi$_2$Se$_3$.
 Olbricht {\it et al.} \cite{Olbrich14prl}, however, pointed out that a
substantial photon drag effect comparable to the photogalvanic effect
has even been observed in 2D materials with vanishing spin-orbit
coupling such as graphene \cite{Glazov14}.
 For (Bi,Sb)Te based TIs, it has been shown that the photon drag effect
can even outweigh the photogalvanic effect under illumination with
THz radiation at large angles of incidence \cite{Plank16prb}.

Kastl {\it et al.} combined the pickup of the photocurrent by
electrical contacts with a time-resolved detection using Au strip
lines and a laser-triggered Auston-switch that allowed a time-of-light
analysis of the photogenerated hot electrons \cite{Kastl15natcomm}.
 They concluded that the helicity-dependent contribution of the
detected photocurrent has in fact its origin in the TSS
because they found that it just travels with a group velocity that is
consistent with the slope of the Dirac cone of Bi$_2$Se$_3$ at the
Fermi energy.

The photon energy in these experiments exceeded the bulk band gap of
the sample by far which prohibits a direct optical excitation within
the Dirac cone close to $E_F$ as has been considered by Junck {\it et
  al.} \cite{Junck13prb}.
 Moreover, Bi$_2$Se$_3$ is intrinsically significantly n-doped and
the TSS is occupied up to energies above the conduction band minimum.
 This raises the question which electronic states at the surface are
involved in the optical excitation and finally carry the detected
photocurrent in these transport measurements.
 Recent experiments on gated (Bi$_{1-x}$Sb$_x$)$_2$Te$_3$ thin films,
which were utilizing a similar experimental scheme as McIver {\it et
  al.}, together with first principal calculations and an analytical model
suggest that the photocurrents are generated by optical
transitions between the TSS and higher lying bulk bands
\cite{Pan17natcomm}.

 A first more detailed insight into the electronic states that are
involved in the optical excitation of Bi$_2$Se$_3$ by 1.5~eV light was
provided by Niesner {\it et al.} \cite{Niesner12prb}, who demonstrated
by 2PPE that the surfaces of Bi$_2$Te$_x$Se$_{3-x}$ compounds,
including Bi$_2$Se$_3$, support a second unoccupied Dirac cone at an
energy of around 1.5~eV above $E_F$.
 Sobota {\it et al.}  \cite{Sobota13prl} showed shortly afterwards that
it can in fact be directly excited by 1.5-eV laser pulses in n-type
Bi$_2$Se$_3$ thin films.
 2PPE experiments by Ketterl {\it et al.} on Bi$_2$Se$_3$ bulk
crystals employed circular polarized pump pulses and an electron
detection in two-dimensional momentum space \cite{Ketterl18prb}.
 They have shown that 1.7-eV pump pulses provide a direct coupling
between the first and second TSS, but found that the observed
threefold-symmetric dichroic signal is independent of the excitation
energy and only reflects the excitation pattern of the initial state.
 They found a small residual asymmetry, which is compatible with an
unidirectional photocurrent, only in the energy region where the first
TSS hybridizes with bulk states. The latter are, however, not symmetry
protected and cannot carry spin-polarized photocurrents.
 This let them conclude that the helicity-dependent photocurrents
excited by near-infrared light in Bi$_2$Se$_3$ do not reflect an
intrinsic property of the TSS.

 Recent 2PPE experiments on Bi$_2$Se$_3$ by Soifer {\it et al.}
\cite{Soifer19prl} showed that 3-eV pump pulses create
helicity-dependent asymmetries of different sign across the unoccupied
spectrum due to different resonant optical transitions, including
transitions into the 2nd TSS, but that not each asymmetric
distribution is associated with a TSS.
 All observed asymmetries decay on a time scale of the cross
correlation between pump and probe pulses and no persistent asymmetry
was observed.

These experiments on Bi$_2$Se$_3$ show that even if the origin of a
photocurrent can be linked to the surface of a 3D TI, it is not {\it a
  priori} connected to the unusual transport properties of the TSS,
which promises long-living currents with only low losses.

\subsection{Helicity-dependent 2PPE on Sb$_2$Te$_3$}

\begin{figure*}
 \includegraphics[width=0.65\textwidth]{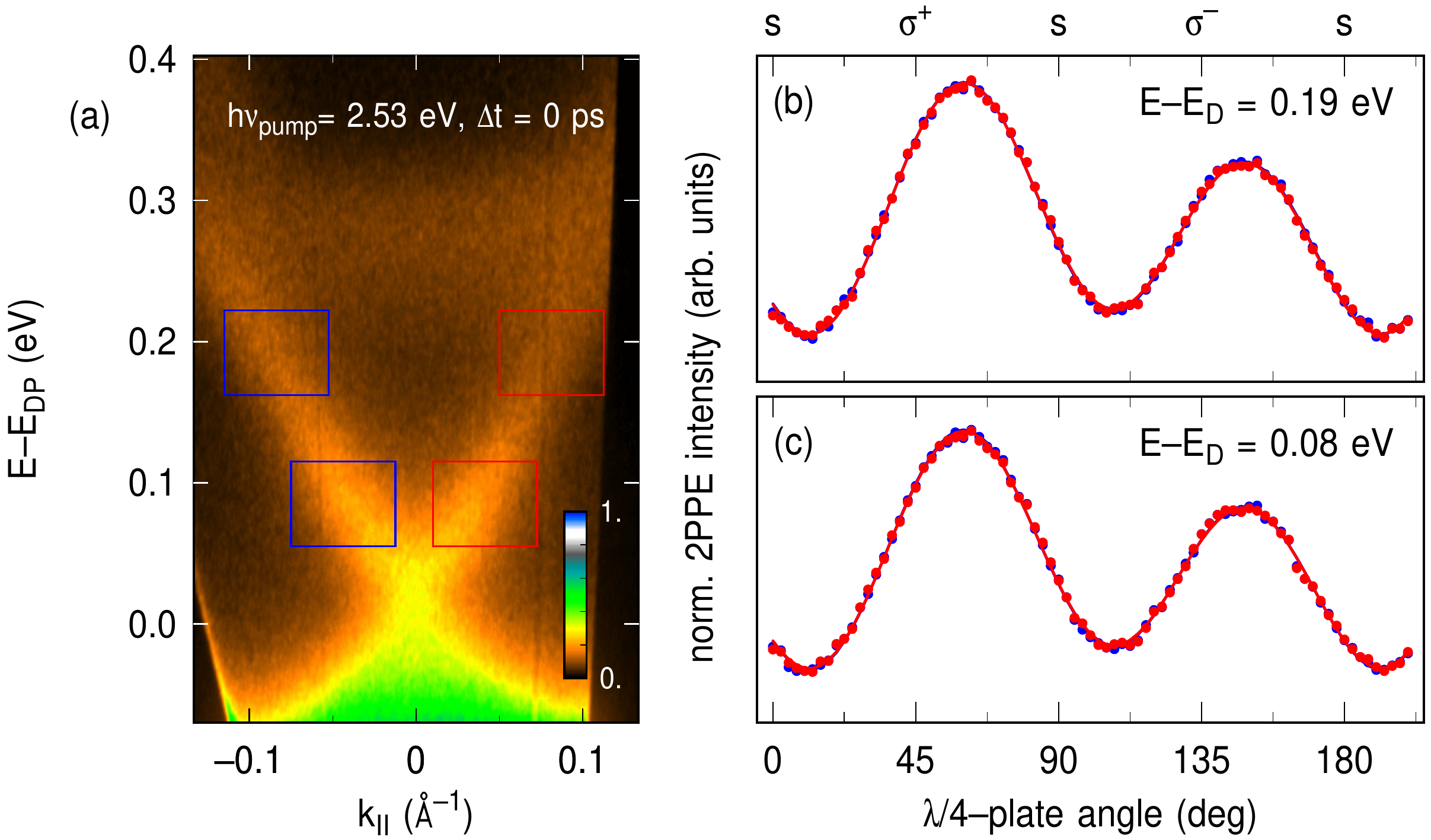}
 \caption{(a): 2PPE spectrum of the TSS of Sb$_2$Te$_3$ at temporal
   overlap between UV probe and visible pump pulses. The electrons
   were detected along the $\bar{\Gamma}$-$\bar{K}$ direction,
   perpendicular to the plane of light incidence. (b) and (c):
   Normalized 2PPE signal at $-k_\parallel$ (blue) and $+k_\parallel$
   (red) as a function of the polarization of the pump pulses for two
   energy regions as indicated by the integration windows in (a),
   respectively. The absence of a phase shift between the data for
   opposite $k_\parallel$ shows that no helicity-dependent
   photocurrent is generated.}
  \label{fig:circular_vis}
\end{figure*}

Motivated by the first experiments of McIver {\it et al.}
on Bi$_2$Se$_3$ \cite{McIver12natnano}, we systematically explored the
possibility to detect a helicity-dependent photocurrent in
Sb$_2$Te$_3$ by 2PPE using pump pulses of different photon energies (0.52-1.03 , 1.55, 2.53 eV).
 For these, and all of experiments presented in the following,  the electrons were
detected in a direction perpendicular to the plane of incidence of
pump and probe pulses in order to exclude the generation of a k-space
asymmetry of the detected photoelectrons due to linear dichroism of
the probe pulses at oblique incidence as has been observed the
first experiments on Sb$_2$Te$_3$ \cite{Reimann14prb}.
 The helicity of the initially s-polarized pump pulses was varied by
rotating a $\lambda/4$-plate which was placed in the pump beam path.
 Exemplarily, Fig.~\ref{fig:circular_vis} shows the variation of the 2PPE intensity
for opposite $k_\parallel$ at two different energies above the Dirac
point upon excitation with 2.53-eV pump pulses.
The plane of light incidence was oriented
along $\bar{\Gamma}$-$\bar{M}$ of the Sb$_2$Te$_3$ surface
Brillouin zone (SBZ) and the photoelectrons were detected
perpendicular to this direction along $\bar{\Gamma}$-$\bar{K}$.
 The 2PPE intensity has been normalized in
Fig.\ref{fig:circular_vis}(b) and (c) because a slight asymmetry is
often observed because of a non perfect alignment of the sample.
 These plots show that the pump helicity changes the population
of the TSS symmetrically but does not introduce a phase shift between opposite
parallel momenta $k_\parallel$.
 This means that the variation of the helicity simply changes the
absorption of the pump pulses because of the different Fresnel
coefficients for the s- and p-polarized components of the electrical
field.
 The absence of a phase shift clearly shows that no helicity-dependent
photocurrent is generated.
 This negative result agrees with the fact that, unlike in the case of
Bi$_2$Se$_3$, a second unoccupied TSS has not been identified for
Sb$_2$Te$_3$.

In contrast, Sanch{\'e}z-Barriga {\it et al.} \cite{Sanchez16prb}, who
performed spin- and time resolved photoemission experiments for
Sb$_2$Te$_3$, reported such a helicity-dependent photocurrent upon
pumping with 1.5 eV and attributed it to a resonant transition from
deeper-lying bulk valence-band states at $\sim 1$~eV below $E_F$ to
the TSS.  Since the sample orientation has been used as in our
experiment, this discrepancy is difficult to understand if the
samples have the same electronics structure.  Compared to the results
of Refs.~\cite{Reimann14prb,Kuroda16prl,Sumida17scirep}, however, the
lifetime of the electrons in the TSS was found to be shorter by more
than a factor of two in Ref. \cite{Sanchez16prb}, although the position
of $E_F$ was similar.

\section{Mid-IR excitation}
\label{sec:mid-IR-excitation}

\subsection{Direct optical excitation of the TSS}
\label{subsec:direct-excitation}

\begin{figure*}
 \includegraphics[width=\textwidth]{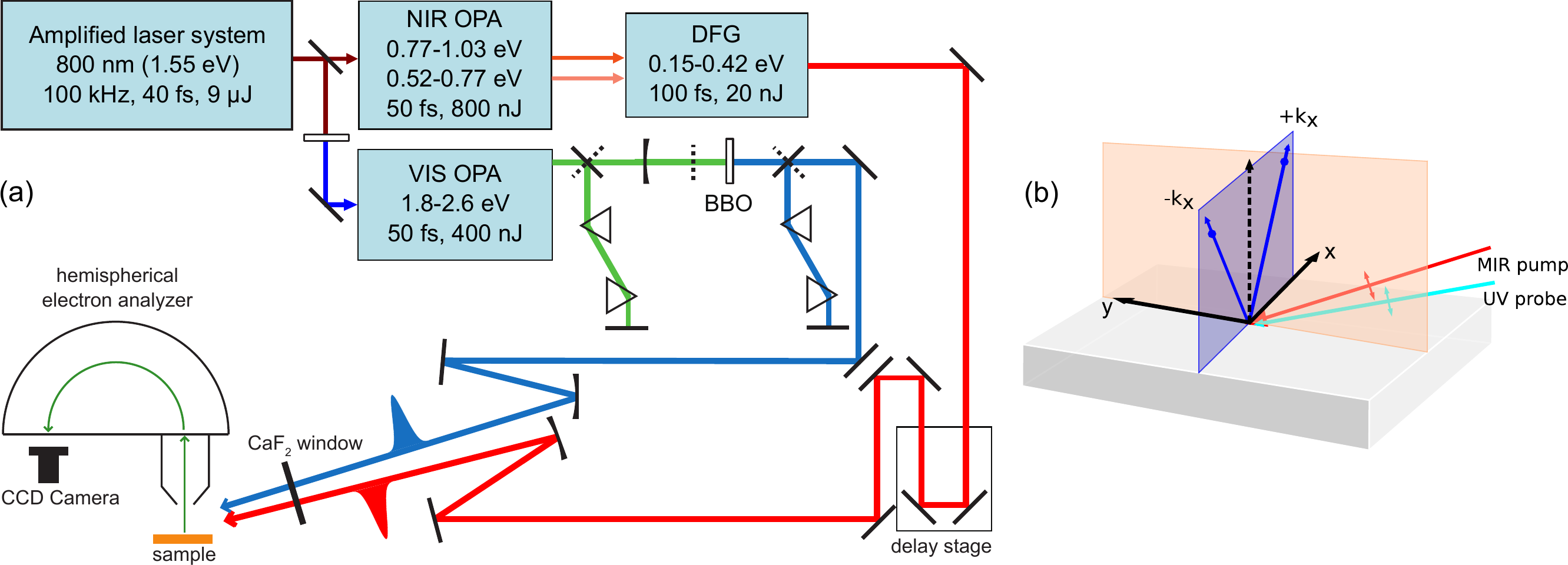}
 \caption{(a) Experimental setup for 2PPE with mid-IR pump pulses and UV probe pulses.
   (b) Experimental geometry. The entrance slit of the hemispherical electron analyzer
     ($k_X$- or $k_\parallel$-direction) is oriented perpendicular to the plane of light incidence.
   }
  \label{fig:mir-setup}
\end{figure*}

 In order to restrict the optical excitation to transitions in the
close vicinity of the TSS, Kuroda {\it et al.} \cite{Kuroda16prl}
developed a 2PPE setup that utilizes ultrashort mid-IR pump pulses
with photon energies that match the bulk band gap of Sb$_2$Te$_3$.
 This was realized by using a laser system that is able to pump two
optical parametric amplifiers (OPA) simultaneously.
A scheme of this setup together with the experimental geometry
is depicted in Fig.~\ref{fig:mir-setup}.
 The frequency doubled output of the primarily used OPA, which is
operated in the visible range, still provides the UV probe
pulses.
 Pump pulses with photon energies of 0.2-0.4~eV were generated by
difference frequency mixing of the signal and idler output of the
second OPA, which is tunable in the near infrared.
 With these low photon energies, a novel direct optical
excitation process of the TSS by a resonant transition from its
occupied into its unoccupied part across the Dirac point
could be revealed \cite{Kuroda16prl}.
 Moreover, it was found that even linearly polarized mid-IR light is able
to produce a strong asymmetric population of the TSS in in k-space.
 By observing the decay of the asymmetric population,
the dynamics of elastic momentum scattering, which is restricted
due to the protection against backscattering, could be investigated \cite{Kuroda16prl}.

\begin{figure*}
 \includegraphics[width=\textwidth]{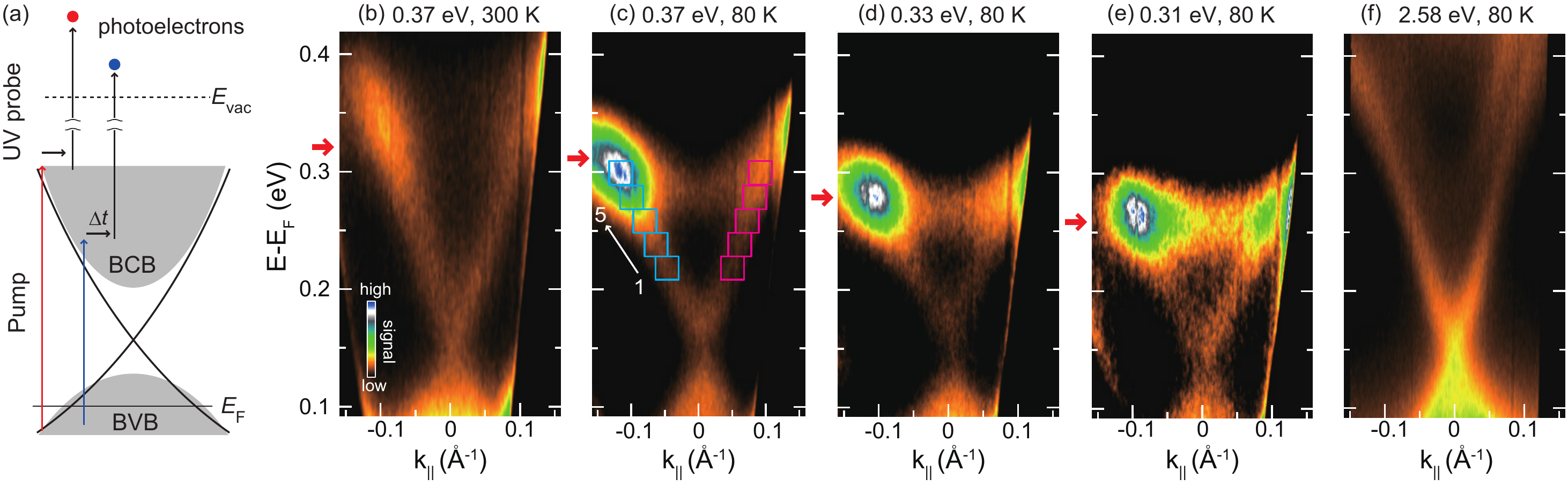}
 \caption{(a) Excitation scheme for the population of the TSS in
   Sb$_2$Te$_3$ with mid-IR pump pulses and subsequent photoemission with
   ultraviolet probe pulses. (b) and (c) Angle-resolved 2PPE spectra
   for 0.37 eV pump and 5.16 eV probe pulses 50 fs after mid-IR
   excitation at 300 K and 80 K, respectively. The electrons
   were detected along a direction close to $\bar{\Gamma}$-$\bar{M}$,
   perpendicular to the plane of light incidence. Red arrows indicate the
   energies of maximum population enhancement in the TSS. (d) and (e)
   show the results acquired at 80 K using 0.35 eV and 0.31 eV pump
   pulses, respectively. (f) shows a spectrum for 2.58 eV pump pulses
   for comparison. Reproduced with permission.\cite{Kuroda16prl}
   Copyright 2016, American Physical Society.}
  \label{fig:Kuroda16prl-fig1}
\end{figure*}

 Figure~\ref{fig:Kuroda16prl-fig1} demonstrates the resonant optical
excitation of the TSS as well as the generation of an asymmetric
population in $k$-space for low photon energies by comparison of a
2PPE spectrum just after visible excitation with spectra after
excitation with mid-IR pump pulses of photon energies between 0.31 eV
and 0.37 eV.
 It shows that only the mid-IR pump pulses generate a strongly enhanced
population of the TSS at a specific energy above the Dirac point (red
arrows) that shifts downwards with decreasing photon energy.
 By evaluating this energy position as a function of the mid-IR photon
energy, it could be revealed that it originates from resonant transitions
between the occupied and unoccupied part of the TSS across the Dirac point
\cite{Kuroda16prl}.
 The 2PPE spectra in Fig.~\ref{fig:Kuroda16prl-fig1} show in addition
that excitation with p-polarized mid-IR pulses also generates a strong
asymmetry between the 2PPE intensity opposite $k_\parallel$.
 Because this asymmetry is not observed for excitation with visible pump pulses
but the same UV probe pulses, linear dichroism of the photoemission probe process
cannot be responsible for this asymmetry but in fact represents an
asymmetric population of the TSS in $k$-space.

\subsection{Photocurrent generation}
\label{subsec:photocurrent-generation}

For a one-dimensional system, a population difference for opposite
$k_\parallel$ can be causally equated with a macroscopic
photocurrent.
 For the two-dimensional TSS of a tetradymite 3D TI, however, the
observation of a population asymmetry along a line in the 2D $k$-space
of the surface is a necessary but not sufficient condition for the
presence of a macroscopic photocurrent because it could also result from a
threefold symmetric excitation of the TSS as has been pointed out by
Ketterl {\it et al.}  \cite{Ketterl18prb}.
In order to answer this question, we have recently employed a novel
hemispherical electron analyzer (Scienta DA30) that is equipped with deflection
plates in the electron lens.
 This makes it possible to acquire energy-momentum ($E$-$k_x$) maps with
the electron momentum $k_x$ oriented along the orientation of the
entrance slit of the hemisphere for varying momentum $k_y$
perpendicular to $k_x$.
 In this way, the full surface band structure can be sequentially mapped in
two-dimensional momentum without moving the sample.
 It should be noted that this method is qualitatively different from
the commonly applied variation of the sample azimuth because it keeps
the direction of the light incidence fixed with respect to the sample
orientation.
 We will show in the following that the application of this technique
reveals that the observed asymmetry of the population for opposite
$k_\parallel$ in the case of mid-IR excitation is in fact associated
with a macroscopic photocurrent in the TSS.

\begin{figure*}
 \includegraphics[width=12cm]{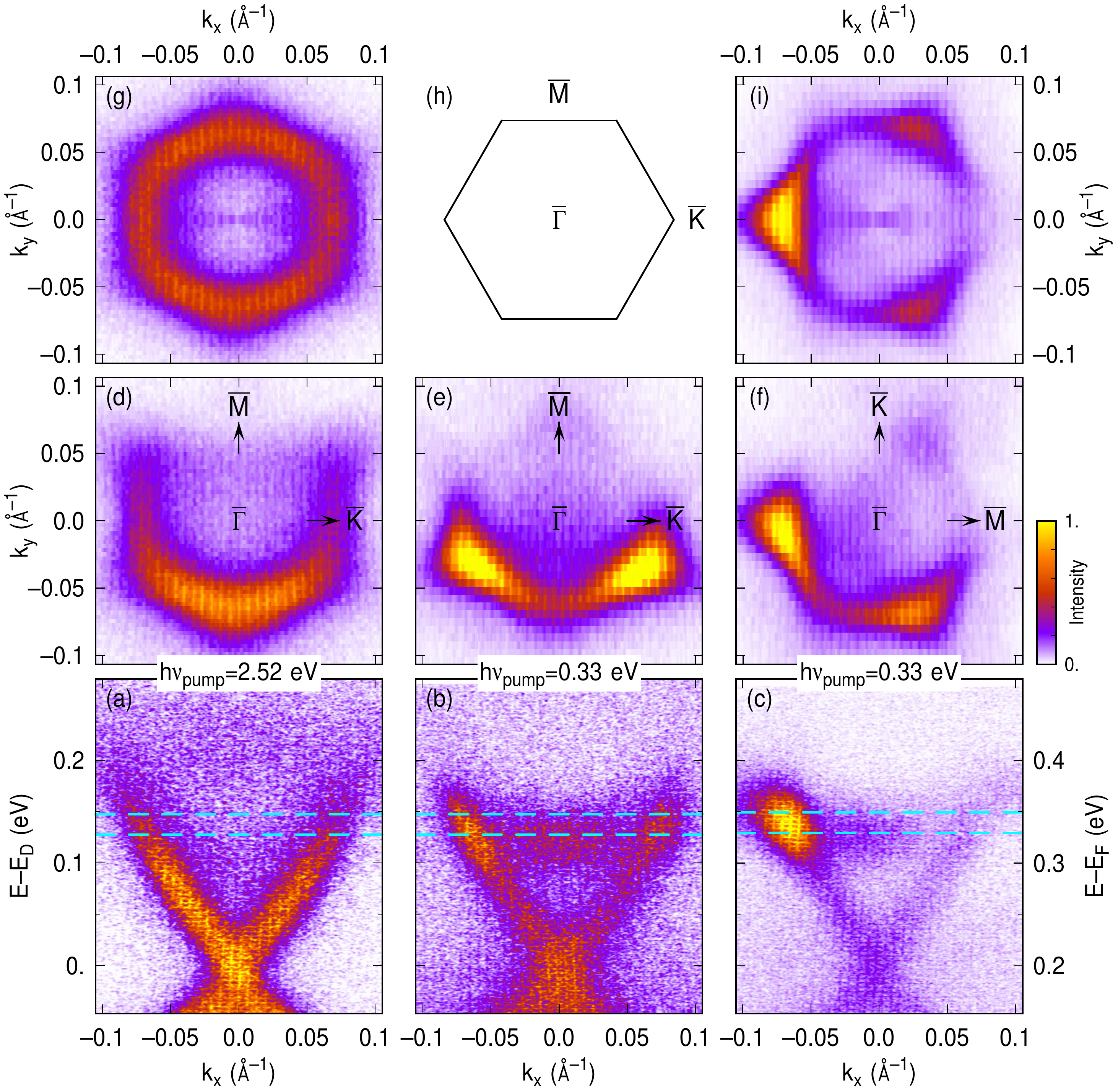}
 \caption{Angle-resolved 2PPE data of the TSS of Sb$_2$Te$_3$ excited
   by visible and by mid-IR pump pulses for different sample
   orientations. (a) $E$-$k_x$ map for 2.52-eV excitation and $k_x$
   along $\bar{\Gamma}$-$\bar{K}$.  (b) for 0.33-eV excitation and
   $k_x$ along $\bar{\Gamma}$-$\bar{K}$.  (c) for 0.33-eV excitation
   and $k_x$ along $\bar{\Gamma}$-$\bar{M}$.  (d)-(f) Corresponding
   $k_x$-$k_y$ maps integrated over energy intervals centered at
   $E-E_D=140$~meV as is depicted by the cyan dashed lines in (a)-(c).
   (g) and (i) show the same cuts as (d) and (f) but with the
   intensity corrected for the matrix element of the probe transition
   and symmetrized by mirroring the data at the mirror axis of the
   surface ($\bar{\Gamma}$-$\bar{M}$-direction). For all data, the
   plane of light incidence is oriented along the $k_y$-direction.
   (h) Surface Brillouin zone of Sb$_2$Te$_3$.}
  \label{fig:scienta}
\end{figure*}

 Figure~\ref{fig:scienta} compares selected cuts of the full
two-dimensional surface band structure of the TSS of Sb$_2$Te$_3$ for
visible and mid-IR excitation and two different sample orientations.
 The lower row shows $E$-$k_x$ maps of the 2PPE intensity as has been
presented above, whereas the middle row shows $k_x$-$k_y$ maps for
which the 2PPE intensity has been integrated over an energy interval
that is centered at the resonantly excited energy for the mid-IR
excitation.

 Although the excitation with visible light results in a homogeneous
population of the full Dirac cone due to sequential filling from
higher-lying states, the 2PPE intensity in the $k_x$-$k_y$ map shown in
Fig.~\ref{fig:scienta}(d) reveals a reduced two-fold symmetry with
a mirror axis along the direction of light incidence ($k_y$-direction).
 The data is therefore symmetric only along the $k_x$-direction as
shown in Fig.~\ref{fig:scienta}(a).
 This asymmetry, however, does not result from an inhomogeneous population of the TSS
 but from the oblique incidence of the p-polarized UV probe pulses
 and is independent of the sample orientation as has been
   tested by azimuthal rotation of the sample by 90$^\circ$ (not shown).
In general, the 2PPE intensity distribution is governed by both the
transient population of the intermediate state excited by the pump
pulses and the sequential photoemission by the probe pulses into the
detected final states.
 In the case of visible excitation, any momentum dependence of the
spectral weight of the 2PPE data can be related to the photoemission
probe process. It depends on the polarization of the probe pulses and
the symmetry of the intermediate state.
 For $p$-polarized probe pulses incident along the $k_y$-direction, a
half moon shaped intensity distribution indicates that the TSS is
dominated by out-of-plane $sp_z$ orbitals with negligible in-plane
contributions \cite{Moser17jelsp}.
 The latter would result in a threefold symmetric pattern.
 The $k_x$-$k_y$ maps can be therefore corrected for the photoemission
probe process by dividing the intensity by $(1-\sin\phi)$ where $\phi$
is the azimuthal angle counting anticlockwise with respect to the
$+k_x$ direction.
 This is, however, not applicable for $\phi$ close to 90$^\circ$ and
this correction is only applied for $k_y<0$ and mirror the data with
respect to the $k_x$ axis.
 The corrected and symmetrized data depicted in Fig.~\ref{fig:scienta}(g)
shows in fact a homogeneous intensity distribution around the Dirac cone and
in particular no kink at the mirror axis.
 Moreover, at this energy above the DP, the warping of the Dirac cone
with a slight flattening of the linear dispersion along
$\bar{\Gamma}$-$\bar{M}$ \cite{Menshc11jetpl} becomes visible.
 The correction of the 2PPE intensity for the impact of the photoemission probe
makes it possible to reveal the actual population in the
intermediate state by this intensity correction also for other pump
photon energies as long as the same photoemission probe is used.
 In the case of an inhomogeneous population in $k$-space, the mirroring
of the data can, however, only be applied if a mirror axis of the
sample surface is oriented perpendicular to the plane of light
incidence.

A mirror axis of the three-fold symmetric surface of Sb$_2$Te$_3$(0001)
is found along the $\bar{\Gamma}$-$\bar{M}$ direction. If this axis
is oriented perpendicular the plane of incidence ($k_y$-direction), the 2PPE
intensity reveals a strong asymmetry with respect to the $k_x$-direction
as can be seen in Fig.~\ref{fig:scienta}(c).
 The corresponding raw and corrected $k_x$-$k_y$ maps shown in
Fig.~\ref{fig:scienta}(f) and (i), respectively, shows that the mid-IR
excitation does not produce a simple two-fold symmetric population of
the TSS but an enhancement of the population at three points within
the Dirac cone along $\bar{\Gamma}$-$\bar{M}$ directions, which
reflects the three-fold symmetry of the Sb$_2$Te$_3$(0001) surface.
Even if the photoemission probe is much less efficient in the
direction of the upper right $\bar{M}$ point, the enhancement is still
faintly visible in the uncorrected data.
 The pattern indicates that the optical excitation is associated with
the Sb-Te bonds which have a three-fold symmetric arrangement in the unit cell
\cite{Glinka15prb}.
 The degree of the population enhancement, however, differs in the
three directions.
 Even the raw data in Fig.~\ref{fig:scienta}(f) clearly shows that it
is much stronger in direction of the left ($\phi=180^\circ$) as
compared to that in the direction to the lower right
($\phi=-60^\circ$).
 Quantitatively, the corresponding 2PPE intensity is larger by 40\%.
This difference is further enhanced by a factor of about 1.9 when the
data is corrected for the photoemission probe as shown in
Fig.~\ref{fig:scienta}(i) but it should be emphasized that
the conclusion on the asymmetric population is not based on this
correction procedure.
 The symmetrized image shown in Fig.~\ref{fig:scienta}(i)
should only serve for a better visualization.

 These results  unambiguously show that the direct excitation by the mid-IR
pulses in fact generates a macroscopic photocurrent in the TSS along the $k_x$
direction.
 Because of the spin texture of the TSS, this photocurrent should be
automatically spin-polarized.

The magnitude and sign of the intensity asymmetry observed along the $k_x$-direction
depends on the azimuthal orientation of the sample with respect to the
plane of light incidence \cite{Kuroda17spie}.
 As shown in Fig.~\ref{fig:scienta}(b), it vanishes if the
$\bar{\Gamma}$-$\bar{M}$ direction is oriented along the plane of
incidence.
 In this case, the excitation pattern in the TSS has a pure three-fold
symmetry and no photocurrent is generated as can be concluded from
Fig.~\ref{fig:scienta}(e).
 Note that symmetrization is not possible in this three-fold
symmetric distribution because the $\bar{\Gamma}$-$\bar{K}$
direction is not a mirror plane.

On the basis of these new results, we take the opportunity to clarify
the following point.
 In Ref.~\cite{Kuroda16prl}, it was reported that the asymmetry for
opposite $k_\parallel$ was observed, if the plane of light incidence
was oriented along $\bar{\Gamma}$-$\bar{M}$ and the electrons were
detected along $\bar{\Gamma}$-$\bar{K}$.
 For these experiments, the sample azimuth was determined by low-energy
electron diffraction (LEED) with an accuracy of better than 5$^\circ$
with respect to the orientation of the entrance slit of the
hemispherical electron analyzer.
 The two-dimensional mapping of the photoelectrons presented in
Fig.~\ref{fig:scienta}, however, makes it possible to determine the
sample orientation with respect to the electron spectrometer
in a direct way with the help of the warping of the Dirac cone.
 This reveals that the asymmetry and therewith the photocurrent is
in fact generated for the opposite sample orientation.
Examination of the $k_x$-$k_y$ map shown in Fig.~\ref{fig:scienta}(e), however,
makes it clear that even a small misalignment of the sample azimuth results
in a strong asymmetry in the $E$-$k_x$ cuts.
We therefore conclude that the limited precision of the azimuth determination
in Ref.~\cite{Kuroda16prl} was responsible for the contrary assignment.

\subsection{Ultrafast decay dynamics of the photocurrent}
\label{subsec:decay-dynamics}

\begin{figure*}
 \includegraphics[width=\textwidth]{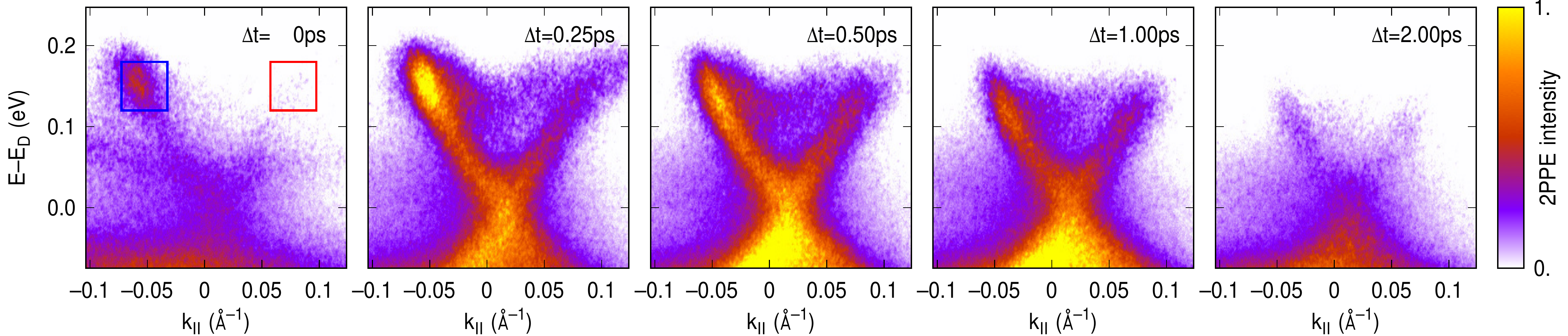}
 \caption{Angle-resolved 2PPE spectra of Sb$_2$Te$_3$ for different
   delays between the mid-IR pump and the UV probe pulses.  The
   overall time resolution (convolution of pump and probe pulses) is
   better than 200~fs (FWHM). The electrons were detected along a
   direction close to $\bar{\Gamma}$-$\bar{M}$, perpendicular to the
   plane of light incidence.}
  \label{fig:Kuroda17spie-movie-v1}
\end{figure*}

The generation of a photocurrent makes it possible to utilize 2PPE
not only for the investigation of inelastic decay but makes it
possible to access also elastic momentum scattering within TSS by observing
the redistribution of the electron distribution in $k$-space as was
demonstrated in Ref.~\cite{Kuroda16prl}.
 In this way, it could be shown that the spin structure of the TSS in fact
imposes strong restrictions on momentum scattering which are reflected
by a slow homogenization of the initially inhomogeneous distribution
of the electrons in momentum space.
 This can be already concluded from a series of 2PPE spectra for
selected delays $\Delta t$ between the mid-IR pump und UV probe pulses
as is compiled in Fig.~\ref{fig:Kuroda17spie-movie-v1}.
 This series shows that the population of the TSS at the resonantly excited
energy has its maximum already at $\Delta t=0.25$~ps, which reflects
the direct optical excitation without delayed filling.
 The filling of the lower part of the Dirac cone, however, is delayed
and peaks at $\Delta t = 0.5$~ps due to sequential inelastic
scattering of the resonantly excited electrons to lower energies.
 At temporal overlap between pump and probe pulses ($\Delta t = 0$), an
additional parabolic band can be observed at $\sim 50$~meV above the
DP.
 This band results from population of the third image-potential state,
which is in fact located close below the vaccum level but appears at a
similar final state energy as the TSS.
 For 2PPE of image-potential states, the role of pump and probe pulses
is reversed and their decay is observed for negative delays of the UV
pulses with respect to the mid-IR pulses.
 For increasing delay, the population of the TSS gradually
decreases but asymmetry between opposite $k_\parallel$ is still
visible even at $\Delta t = 2$~ps. This already shows that the randomization
of the TSS population in $k$-space proceeds on a slower time scale as compared to the
inelastic decay.

\begin{figure}
 \includegraphics[width=0.8\columnwidth]{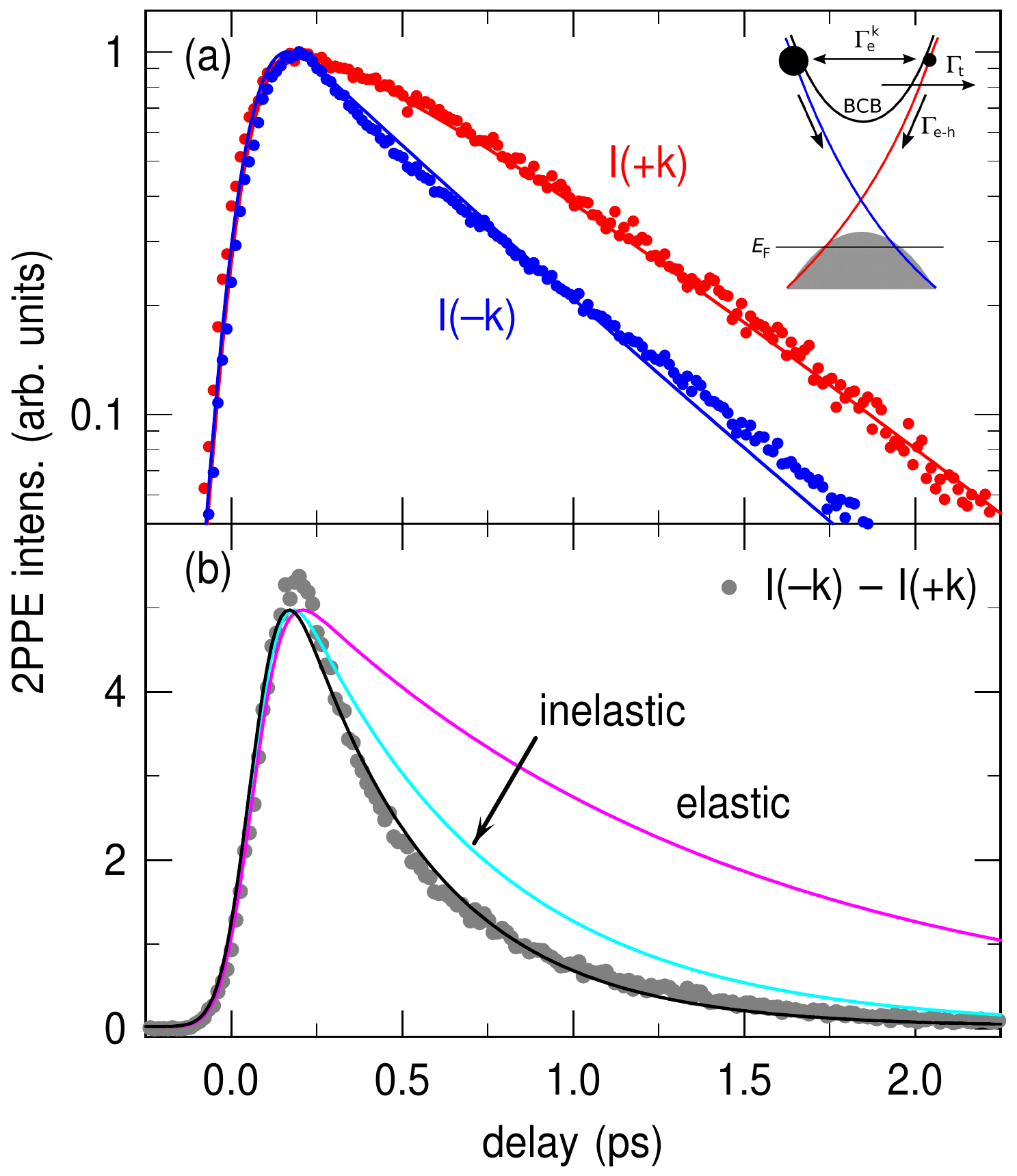}
 \caption{(a) Normalized 2PPE intensity of the transient population in
   the Dirac cone of Sb$_2$Te$_3$ at opposite parallel momenta
   $k_\parallel$ (data points). Solid lines are fits to the data (see
   text). The two data sets have been obtained by integration of the
   2PPE intensity over blue and red rectangles such as shown in
   Fig.~\ref{fig:Kuroda17spie-movie-v1} but for data published in
   Ref.~\cite{Kuroda16prl}. The inset shows a scheme of the
   rate-equation model for fitting the data. (b) Intensity difference
   $\Delta I = I_{-k_\parallel}-I_{+k_\parallel}$ (grey dots). The
   black line shows a fit that considers elastic momentum scattering
   and inelastic decay. The magenta and cyan lines show these
   contributions separately. Reproduced with permission.\cite{Kuroda17spie}
   Copyright 2017, Society of Photo‑Optical
   Instrumentation Engineers (SPIE)}
  \label{fig:Kuroda17spie-fig5}
\end{figure}

 Figure~\ref{fig:Kuroda17spie-fig5} (a) shows the 2PPE intensity
integrated over regions as are depicted by the blue and red rectangles
shown in the most left graph of Fig.~\ref{fig:Kuroda17spie-movie-v1}
at the direct excitation energy but for comparable data published in
Ref.~\cite{Kuroda16prl}.
 It reveals a distinct different decay dynamics for opposite
$k_\parallel$.
 At $-k_\parallel$ (blue curve), the decay is initially faster and
becomes slower for delays larger that $\sim 750$~fs.
 In contrast to this, the 2PPE intensity at $+k_\parallel$ is initially
slower.
 For larger delays, the decay rate matches the one observed
for $-k_\parallel$. Both curves proceed with the same slope in the
semi-logarithmic plot.
 The initially faster delay at $-k_\parallel$ and the delay response
at $+k_\parallel$ can be explained by the transfer of the electrons
from $-k_\parallel$ to $+k_\parallel$ due to momentum scattering along a
circular constant energy cut of the Dirac cone.
 Simultaneously, the population at both $k_\parallel$ decay by inelastic
scattering with the same rate, which explains the common decay for larger delays.

In Ref.~\cite{Kuroda16prl}, it has been shown that the decay dynamics at
opposite $k_\parallel$ can be decomposed into inelastic and elastic
decay with the help of a simple rate-equation model, which is depicted
in the inset of Fig.~\ref{fig:Kuroda17spie-fig5} (a).  It considers two
momentum-independent effective decay rates $\Gamma_i$ and $\Gamma_e^k$.
$\Gamma_i$ describes the decay out of the TSS that consists of true
inelastic decay to lower-lying electronic states ($\Gamma_{e-h}$) as
well as quasi-elastic interband scattering into the bulk conduction
band ($\Gamma_t$). The latter is possible because the energy of the
direct optical excitation is close to the conduction band minimum.
$\Gamma_e^k$ describes the rate of population exchange between
opposite parallel momenta. Because 180$^\circ$ momentum scattering in
a single event should be forbidden due to the protection of the TSS
against backscattering, it represents an effective rate for in fact multiple
scattering events along the two-dimensional Dirac cone.
By assuming that the 2PPE intensities $I_{\pm k_\parallel}$ at opposite parallel
momenta are proportional to the respective populations $n_{\pm k_\parallel}$ of the TSS,
the two rate equations for $I_{+k_\parallel}$ and $I_{-k_\parallel}$ can be written as
\begin{equation}
\frac{dI_{\pm k_\parallel}}{dt}=P_{\pm
  k_\parallel}{\delta}(t)\pm\Gamma_e^k\Delta I-\Gamma_iI_{\pm
  k_\parallel}.
\label{eq:rate_intensity}
\end{equation}
Here, $P_{\pm k_\parallel}$ denotes the different excitation
probabilities at opposite $k_\parallel$, $\delta(t)$ the temporal
profile of the Gaussian shaped mid-IR pulses, and $\Delta I =
I_{-k_\parallel}-I_{+k_\parallel}$ being the intensity difference
between $\pm k_\parallel$.  Both equations can be combined into a
single rate equation for $\Delta I$
\begin{equation}
\frac{d{\Delta}I}{dt}=(P_{-k_\parallel}-P_{+k_\parallel}){\delta}(t)-({2\Gamma^{\rm
    k}_{\rm e}+\Gamma_{\rm i}}){\Delta}I,
\label{eq:rate_current}
\end{equation}
which describes the dynamics of the photocurrent that decays
expontially with a time constant $\tau_c=1/({2\Gamma^{\rm k}_{\rm
    e}+\Gamma_{\rm i}})$.  The best fits of the data, which are shown
by the solid lines in Fig.~\ref{fig:Kuroda17spie-fig5}, yield
$\tau_{\rm c}$=0.42~ps, $\tau_{\rm i}=1/\Gamma_{\rm i}=0.6$~ps and
$\tau_{\rm e}=1/\Gamma^{\rm k}_{\rm e}=2.5$~ps.

The characteristic time for elastic momentum scattering $\tau_{\rm e}$
is therefore considerably longer as compared to the inelastic decay
time $\tau_{\rm i}$.
 It is also much longer if compared, e.g. with
electrons in surface states of well-prepared noble metal surfaces
\cite{Gudde07sci,Fauster07pss,Marks11prb2}.  By considering that the
electrons in the TSS move with a Fermi velocity of $v_F=3$~\AA/fs,
which corresponds to the slope of the Dirac cone of the TSS, an
elastic scattering time of $\tau_{\rm e}=2.5$~ps corresponds to a mean
free path of $\lambda_e=v_F\tau_{\rm e}=0.75$~$\mu$m.
 This is almost two orders of magnitude larger as compared to the
typical mean distance between defects of less than 100~\AA\ on such
cleaved surface as has been reported by scanning tunneling microcopy
(STM) \cite{Jiang12prl} and represents a manifestation of the
suppression of backscattering scattering within the TSS.

\subsection{Polarization control of the photocurrent}
\label{subsec:polarization-control}

\begin{figure*}
 \includegraphics[width=17cm]{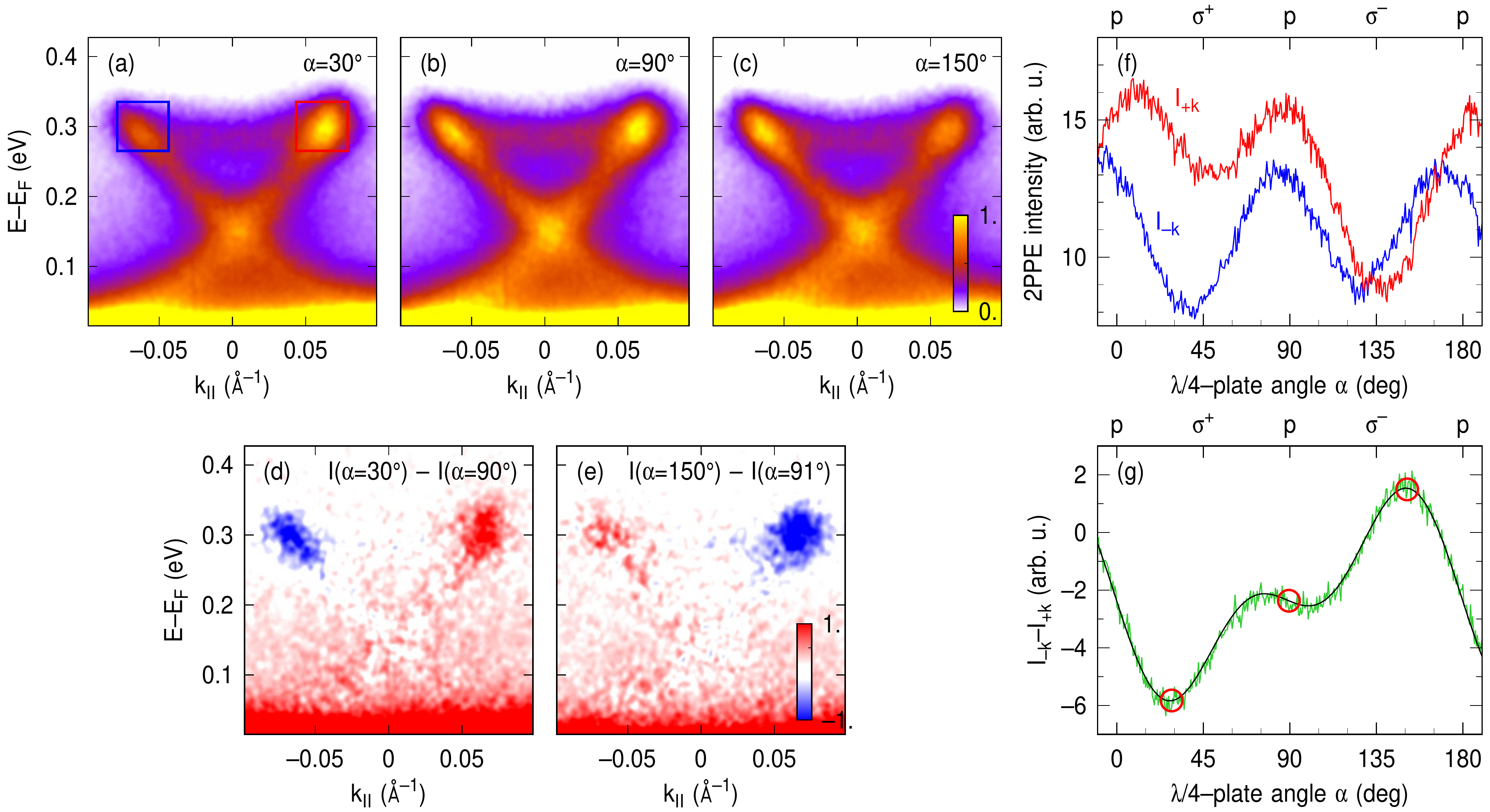}
 \caption{Helicity control of the photocurrent. (a)-(c) Angle-resolved
   2PPE spectra of Sb$_2$Te$_3$ for three different orientations of a
   $\lambda/4$-waveplate in the mid-IR pump beam. The
   $\bar{\Gamma}$-$\bar{M}$ direction of the sample was oriented along
   the plane of light incidence. Electrons were collected allong
   $\bar{\Gamma}$-$\bar{K}$. (d) Intensity difference between spectra (b) and
   (a). (e) Difference between (c) and (b). (f) 2PPE intensity for
   opposite $k_\parallel$ as a function of $\lambda/4$-waveplate angle
   integrated over the blue and red square depicted in (a). (g)
   Difference $(I_{-k_\parallel}-I_{+k_\parallel})$ of the intensities
   plotted in (f).  The solid black line show a fit through the data. The
   three red circles mark the rotation angles used for (a)-(c),
   respectively.}
  \label{fig:control}
\end{figure*}

Although the spin texture of the TSS suggests that a photocurrent,
i.e. an asymmetry of the population for opposite $k_\parallel$, should
be controllable by optical excitation of different helicity
\cite{Hosur11prb}, the experiments presented in sections
\ref{subsec:direct-excitation}-\ref{subsec:photocurrent-generation} have shown
that a strong photocurrent can be already generated by linear
polarized mid-IR pulses if the $\bar{\Gamma}$-$\bar{K}$ direction of
the sample is oriented along the plane of light incidence.
 In Ref.~\cite{Kuroda17prb}, it has been shown that a variation of the
light helicity by introducing a rotable $\lambda/4$-waveplate in the
pump beam results only in a small change of the observed asymmetry of
the intensity for opposite $k_\parallel$ of less than 10\%.

 Here, we show by novel data that the magnitude as well as the sign of the asymmetry
can be controlled by the pump helicity if the $\bar{\Gamma}$-$\bar{M}$
direction of the sample is oriented along the plane of light
incidence.
 For this sample orientation, no photocurrent is generated by
p-polarized mid-IR pulses and no asymmetry is observed along the
$\bar{\Gamma}$-$\bar{K}$ direction as shown in
Figs.~\ref{fig:scienta}(b) and (e).
 Figure \ref{fig:control} demonstrates the polarization control of the
photocurrent in Sb$_2$Te$_3$ in this case with
Fig.~\ref{fig:control}(a)-(c) showing angle-resolved 2PPE spectra for
three selected orientations $\alpha$ of a $\lambda/4$-waveplate that
was introduced in the mid-IR pump beam.
 The orientation $\alpha=90^\circ$ used for Figure~\ref{fig:control}(b)
corresponds to p-polarized mid-IR pulses and results in an almost
vanishing intensity difference for opposite $k_\parallel$.
 The small residual asymmetry can be attributed to a not perfect orientation
of the sample azimuth.
 Figure~\ref{fig:control}(a) and (c) show that the magnitude of the asymmetry
can be enhanced as well as its sign can be switched by changing of the waveplate
angle by $-60^\circ$ and $+60^\circ$, respectively.
 The control of the asymmetry is most pronounced at the resonantly
excited energy as shown in Figs.~\ref{fig:control}(d) and (e) which
show difference plots of Figs.~\ref{fig:control}(a) and (c) with
respect to Fig.~\ref{fig:control}(b).
 These two figures show that rotation of the $\lambda/4$-waveplate by
$-60^\circ$($+60^\circ$) enhances the intensity at
$+k_\parallel$($-k_\parallel$) and simultaneously reduces it at
the opposite parallel momentum.
 These two orientations of the $\lambda/4$-waveplate induce the most
strongest change of the asymmetry as can be seen in
Fig.~\ref{fig:control}(g) where the intensity difference for opposite
$k_\parallel$ at the resonantly excited energy is plotted as a
function of $\alpha$.
Following the analysis of McIver {\it et al.} \cite{McIver12natnano}, it
can be decomposed into contributions of the circular and linear photogalvanic effect
by fitting the asymmetry as follows,
\begin{equation}
  I_{-k_\parallel}(\alpha)-I_{+k_\parallel}(\alpha)=C\sin2\alpha+L_1\sin4\alpha+L_2\cos4\alpha+D.
\end{equation}
Here, $C$ describes the helicity-dependent circular photogalvanic effect.
 The coefficient $L_1$ has been attributed to helicity-independent
linear photogalvanic effect and $L_2$ to a modulation of the
absorptivity \cite{McIver12natnano,Kastl15natcomm}. The parameter $D$
considers a polarization-independent background that results here from
the non-perfect azimuthal alignment of the sample.
 The fit shown in Fig.~\ref{fig:control}(g) yields
$C=-2.56(2)$, $L_1=-1.68(2)$, $L_2=-0.16(2)$, and $D=-2.92(1)$.
 It should be noted that the data can be equally well fitted by omitting
$L_2$ but allowing instead a small misalignment of the $\lambda/4$-waveplate
from the nominal orientation $\alpha'=\alpha+\alpha_0$ with $\alpha_0=0.5(1)^\circ$.
 The contribution of the circular photogalvanic effect to the photocurrent
is in both cases for these data almost twice as large as compared to
the linear photogalvanic effect.
 Due to the contribution of the linear photogalvanic effect, the
maximum asymmetry is not reached for left- and right-circular pump
light ($\alpha=45^\circ$ and $\alpha=135^\circ$, respectively) but
still symmetrically with respect to $\alpha=90^\circ$ for elliptical
polarized light at $\alpha=60^\circ$ and $\alpha=150^\circ$,
respectively.

 In summary, these results demonstrate that full optical control of the photocurrent
by the light helicity is possible for this particular sample
orientation but the magnitude of the helicity-dependent asymmetry is small.
 This is in line with the observations of Ketterl {\it et al.}
\cite{Ketterl18prb}, who found for the same sample orientation also
only a small helicity-dependent asymmetry for excitation of
Bi$_2$Se$_33$ with 1.7-eV pulses.
 As shown in section \ref{subsec:photocurrent-generation}, a much larger
photocurrent can be generated along the $\bar{\Gamma}$-$\bar{M}$
direction by using linear-polarized mid-IR pulses incident along the
$\bar{\Gamma}$-$\bar{K}$ direction.

\section{THz-driven TSS currents}
\label{sec:thz}

\begin{figure}
 \includegraphics[width=\columnwidth]{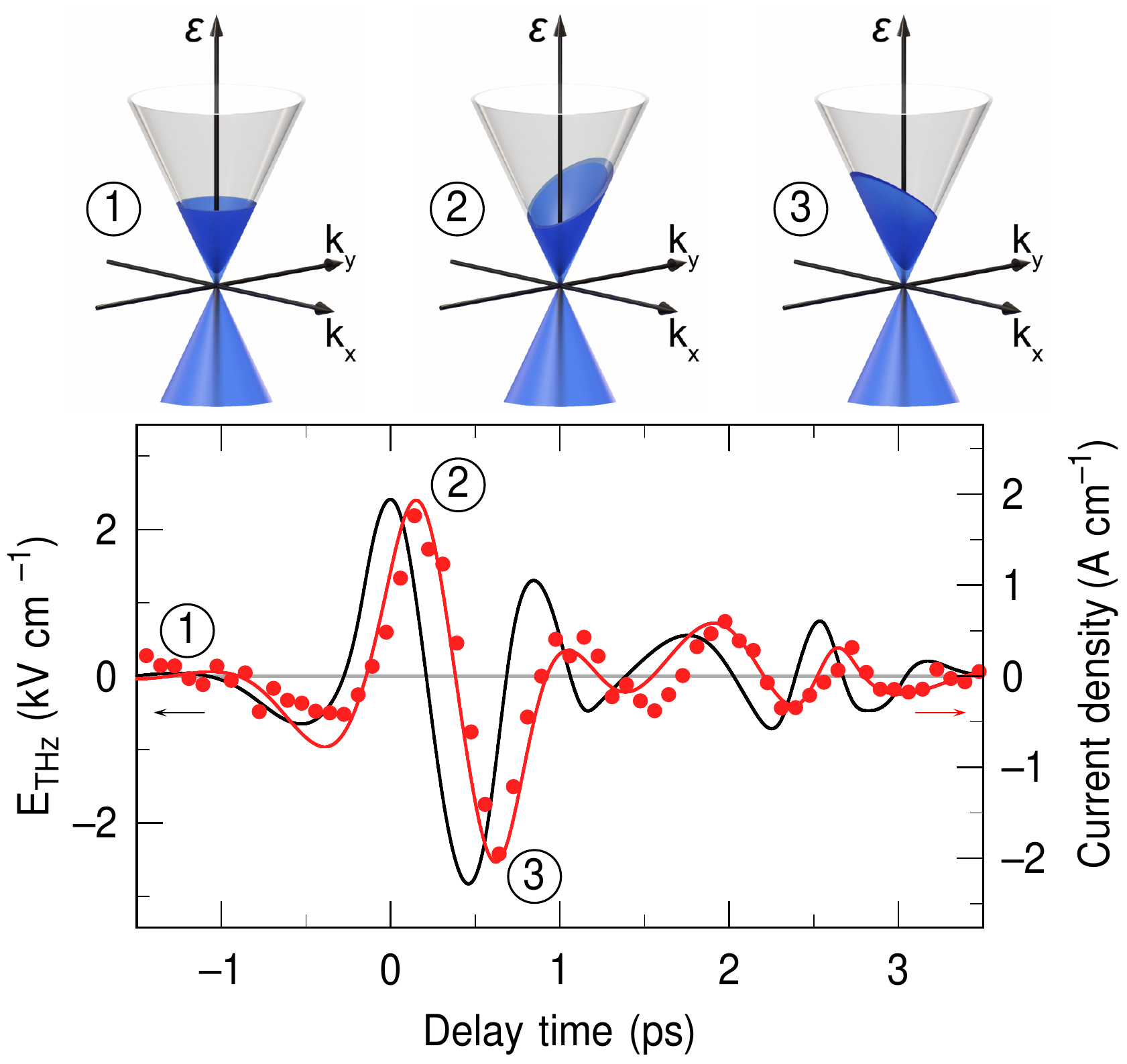}
 \caption{Temporal evolution of the current density in the TSS of
   Bi$_2$Te$_3$ induced by a strong THz electric field pulse with a
   center frequency of 1~THz (data points). The red solid curve shows
   a simulation of the current dynamics using an elastic momentum
   scattering time of 1~ps and an electric field transient that has
   been determined {\it in situ} and that is plotted as black solid
   line (see Ref.~\cite{Reimann18nat} for details). The sketches
   illustrate the acceleration of the Dirac fermions back and forth in
   the topological surface for three selected delays before the
   arrival of the THz transient (1), at the time of maximum
   acceleration in $+k_y$-direction (2), and at the time of maximum
   acceleration in the opposite direction (3). Adapted with permission.\cite{Reimann18nat}
   Copyright 2018, Springer Nature.}
  \label{fig:thz-ARPES}
\end{figure}

The mid-IR excitation scheme makes it possible to generate persistent
photocurrents in the TSS by an optical interband excitation.
 Its combination with time- and angle-resolved 2PPE allows for a very
detailed investigation of the scattering processes of optically
excited electrons in the initially unoccupied TSS of Sb$_2$Te$_3$ and
could verify their large ballistic mean free path.
 In a real device application, however, the electron transport will be
induced by electric fields through intraband acceleration and proceeds
close to the Fermi level.
 Therefore, the most direct way to investigate electron transport on a
microscopic level is to observe how an electric field accelerates
electrons close to E$_F$.

 Recently, such such an experiment has been realized by a combination of
THz excitation with time-resolved ARPES \cite{Reimann18nat}.
 This approach makes it possible to observe directly on a subcycle time scale
how the carrier wave of a terahertz light pulse accelerates Dirac
fermions in the TSS.
 While terahertz streaking of photo-emitted electrons in the vacuum
traces the electromagnetic field at the surface, the acceleration of
the electron in the TSS leads to a strong redistribution of electrons
in momentum space.
 This has been demonstrated for Bi$_2$Te$_3$ where the Fermi level is
located well above the Dirac point but below the conduction band
minimum at the $\bar{\Gamma}$-point.

 In contrast to the mid-IR excitation, no persistent photocurrent is
generated in this experiment because the net acceleration integrated
over the complete transient of the electric field vanishes.
 The subcycle observation of the redistribution of the Fermi-Dirac
distribution, however, makes it possible to investigate the interplay
of back and forth acceleration by the electric field transient and the
scattering of the electrons in the TSS just around the Fermi level.

Figure~\ref{fig:thz-ARPES} summarizes the main results of this work.
The sketches illustrate the back and forth acceleration of the electrons in the
TSS of Bi$_2$Te$_3$ which occupy the TSS up to energies of 200~meV above $E_F$.
 The black solid line depicts the time-evolution of the s-polarized
electric field with a center frequency of 1~THz. It has been
determined {\it in situ} from the momentum streaking of the electrons
in front of the surface.
 Because of the continuity of the s-polarized electric field component
across the interface between vacuum and the sample, this just corresponds
to the field that accelerates the electrons in the TSS along the surface.
 The data points show the transient current density that has been
calculated from asymmetry of the photoelectron spectra at opposite
$k_\parallel$ and the electron density in the TSS.
 The time lag of the current density with respect to the electric field
already demonstrates that the electrons sample energy and momentum
during the acceleration without much losses on the time scale of the
period of the THz transient.
The inertia-free surface currents are protected by spin-momentum
locking and reach peak densities as large as 2 A/cm.
 They travel with the Fermi velocity of 4.1~\AA/fs or 410~nm/ps.
 Comparison of the measured current with the results of a semiclassical
Boltzmann model depicted as red solid line shows that the relevant
scattering times of the electrons carrying the current amount to at
least 1 ps.
 This is a comparable to the results on Sb$_2$Te$_3$ and demonstrates
that 3D topological insulators have in fact the potential to build
ultrafast low-loss electronic devices driven by light waves.

\section{Conclusions and outlook}

The work presented here shows that time- and angle-resolved 2PPE is a
versatile tool to selectively observe the optical generation of
photocurrents and their decay in the TSS of 3D TIs.  It allows to
disentangle the decay of the photocurrent by elastic and inelastic
electron scattering and to verify the exceptional transport properties
in the TSS due to its particular spin texture.  This applies in
particular for the excitation with ultrashort low-energy light pulses in the
mid-IR range, which have shown to permit a direct optical excitation
of the TSS of Sb$_2$Te$_3$ across the Dirac point.  Surprisingly, even
linear polarized mid-IR excitation is able to generate a strong
photocurrent in the TSS while the helicity-dependence of the
excitation is found to be small. For certain sample orientations,
however, the latter provides a full optical control of the magnitude
and sign of the photocurrent.  The mapping of the surface band
structure in two dimensions has shown to be important for the
distinction between population asymmetries along one direction that
simply result from a threefold symmetric excitation and those which
are representative for an actual macroscopic photocurrent.
 2D-mapping of electron redistribution within the full Dirac cone opens
the possibility to access the ultrafast dynamics of optically as well
as THz induced photocurrents in unprecedented detail.
 It makes it possible to observe momentum scattering of arbitrary
scattering angle and to investigate the effect of the warping of the
Dirac cone on the electron and photocurrent dynamics.
 It will furthermore enable to study how defects introduced
by nonmagnetic or magnetic dopants affect not only the band structure
of the TSS but also inelastic and elastic momentum scattering.
 This is of great interest because doping of 3D TIs by nonmagnetic
elements is typically used for tuning the Fermi level while doping
with magnetic elements is one route to realize the quantum anomalous
Hall effect \cite{Tokura19natrev}.

\medskip
\noindent
\textbf{Acknowledgements} \par 

We thank R. Reimann, K. Kuroda, E. V. Chulkov, K. A. Kokh,
O. E. Tereshchenko, A. Kimura, S. Schlauderer, C. P. Schmid,
F. Langer, S. Baierl, C. Lange, and R. Huber for a fruitful and
pleasant collaboration in this work.  We gratefully acknowledge
funding by the Deutsche Forschungsgemeinschaft (DFG, German Research
Foundation) through grant numbers HO 2295/7 (SPP 1666), GU 495/2 and
project ID 223848855-SFB~1083.

\medskip

%



\begin{thebibliography}{10}
\providecommand{\url}[1]{\texttt{#1}}
\providecommand{\urlprefix}{URL }

\bibitem{Fu07prl}
L.~Fu, C.~L. Kane, E.~J. Mele,
\newblock \emph{Phys. Rev. Lett.} \textbf{{2007}}, \emph{{98}}, {10}.

\bibitem{Xia09natphys}
Y.~Xia, D.~Qian, D.~Hsieh, L.~Wray, A.~Pal, H.~Lin, A.~Bansil, D.~Grauer, Y.~S.
  Hor, R.~J. Cava, M.~Z. Hasan,
\newblock \emph{Nat. Phys.} \textbf{2009}, \emph{5} 398.

\bibitem{Zhang09natphys}
H.~J. Zhang, C.~X. Liu, X.~L. Qi, X.~Dai, Z.~Fang, S.~C. Zhang,
\newblock \emph{Nat. Phys.} \textbf{2009}, \emph{5} 438.

\bibitem{Hasan10rmp}
M.~Z. Hasan, C.~L. Kane,
\newblock \emph{Rev. Mod. Phys.} \textbf{2010}, \emph{82} 3045.

\bibitem{Hsieh09nat}
D.~Hsieh, Y.~Xia, D.~Qian, L.~Wray, J.~H. Dil, F.~Meier, J.~Osterwalder,
  L.~Patthey, J.~G. Checkelsky, N.~P. Ong, A.~V. Fedorov, H.~Lin, A.~Bansil,
  D.~Grauer, Y.~S. Hor, R.~J. Cava, M.~Z. Hasan,
\newblock \emph{Nature} \textbf{2009}, \emph{460} 1101.

\bibitem{Gonzales96prl}
J.~Gonz{\'a}lez, F.~Guinea, M.~Vozmediano,
\newblock \emph{Phys. Rev. Lett.} \textbf{{1996}}, \emph{{77}}, {17} {3589}.

\bibitem{Roushan09nat}
P.~Roushan, J.~Seo, C.~V. Parker, Y.~S. Hor, D.~Hsieh, D.~Qian, A.~Richardella,
  M.~Z. Hasan, R.~J. Cava, A.~Yazdani,
\newblock \emph{{Nature}} \textbf{{2009}}, \emph{{460}}, {7259} {1106}.

\bibitem{Fu07prb}
L.~Fu, C.~L. Kane,
\newblock \emph{Phys. Rev. B} \textbf{{2007}}, \emph{{76}}, {4}.

\bibitem{Navratil04jssc}
J.~Navr{\'a}til, J.~Hor{\'a}k, T.~Plech{\'a}\v{c}ek, S.~Kamba, P.~Lo\v{s}t'{\'a}k, J.~Dyck, W.~Chen,
  C.~Uher,
\newblock \emph{{J. Solid State Chem.}} \textbf{{2004}}, \emph{{177}}, {4-5}
  {1704}.

\bibitem{Hor09prb}
Y.~S. Hor, A.~Richardella, P.~Roushan, Y.~Xia, J.~G. Checkelsky, A.~Yazdani,
  M.~Z. Hasan, N.~P. Ong, R.~J. Cava,
\newblock \emph{Phys. Rev. B} \textbf{2009}, \emph{79} 195208.

\bibitem{Chen09sci2}
Y.~L. Chen, J.~G. Analytis, J.~H. Chu, Z.~K. Liu, S.~K. Mo, X.~L. Qi, H.~J.
  Zhang, D.~H. Lu, X.~Dai, Z.~Fang, S.~C. Zhang, I.~R. Fisher, Z.~Hussain,
  Z.~X. Shen,
\newblock \emph{Science} \textbf{2009}, \emph{325} 178.

\bibitem{Hsieh08nat}
D.~Hsieh, D.~Qian, L.~Wray, Y.~Xia, Y.~S. Hor, R.~J. Cava, M.~Z. Hasan,
\newblock \emph{{Nature}} \textbf{{2008}}, \emph{{452}}, {7190} {970}.

\bibitem{Dupont95prl}
E.~Dupont, P.~B. Corkum, H.~C. Liu, M.~Buchanan, Z.~R. Wasilewski,
\newblock \emph{Phys. Rev. Lett.} \textbf{1995}, \emph{74} 3596.

\bibitem{Atanasov96prl}
R.~Atanasov, A.~Haché, J.~L.~P. Hughes, H.~M. van Driel, J.~E. Sipe,
\newblock \emph{Phys. Rev. Lett.} \textbf{1996}, \emph{76} 1703.

\bibitem{Hache97prl}
A.~Haché, Y.~Kostoulas, R.~Atanasov, J.~L.~P. Hughes, J.~E. Sipe, H.~M. van
  Driel,
\newblock \emph{Phys. Rev. Lett.} \textbf{1997}, \emph{78} 306.

\bibitem{Stevens03prl}
M.~J. Stevens, A.~L. Smirl, R.~D.~R. Bhat, A.~Najmaie, J.~E. Sipe, H.~M. van
  Driel,
\newblock \emph{Phys. Rev. Lett.} \textbf{2003}, \emph{90} 136603.

\bibitem{Hubner03prl}
J.~Hübner, W.~W. Rühle, M.~Klude, D.~Hommel, R.~D.~R. Bhat, J.~E. Sipe, H.~M.
  van Driel,
\newblock \emph{Phys. Rev. Lett.} \textbf{2003}, \emph{90} 216601.

\bibitem{Gudde07sci}
J.~G{\"u}dde, M.~Rohleder, T.~Meier, S.~W. Koch, U.~H{\"o}fer,
\newblock \emph{Science} \textbf{2007}, \emph{318} 1287.

\bibitem{Bas15apl}
D.~A. Bas, K.~Vargas-Velez, S.~Babakiray, T.~A. Johnson, P.~Borisov, T.~D. Stanescu,
  D.~Lederman, A.~D. Bristow,
  \newblock \emph{Appl. Phys. Lett.} \textbf{2015}, \emph{106} 041109.

\bibitem{Bas16oe}
D.~A. Bas, R.~A. Muniz, S.~Babakiray, D.~Lederman, J.~E. Sipe, A.~D. Bristow,
\newblock \emph{Opt. Express} \textbf{2016}, \emph{24} 23583.

\bibitem{Aversa95prb}
C.~Aversa, J.~E. Sipe,
\newblock \emph{Phys. Rev. B} \textbf{1995}, \emph{52} 14636.

\bibitem{Muniz14prb}
R.~A. Muniz, J.~E. Sipe,
\newblock \emph{Phys. Rev. B} \textbf{{2014}}, \emph{{89}}, {20}.

\bibitem{Ganichev03jp}
S.~D. Ganichev, W.~Prettl,
\newblock \emph{J. Phys.-Condens. Mat.} \textbf{2003}, \emph{15} R935.

\bibitem{Ivchenko05}
I.~L. Ivchenko,
\newblock \emph{{O}ptical spectroscopy of semiconductor nanostructures},
\newblock Alpha Science Int’l Ltd., Harrow, \textbf{2005}.

\bibitem{Ganichev06}
S.~D. Ganichev, W.~Prettl,
\newblock \emph{Intense Terahertz Excitation of Semiconductors},
\newblock Oxford University Press, Oxford, \textbf{2006}.

\bibitem{Zhang09nat}
Y.~B. Zhang, T.~T. Tang, C.~Girit, Z.~Hao, M.~C. Martin, A.~Zettl, M.~F.
  Crommie, Y.~R. Shen, F.~Wang,
\newblock \emph{Nature} \textbf{2009}, \emph{459} 820.

\bibitem{Hosur11prb}
P.~Hosur,
\newblock \emph{Phys. Rev. B} \textbf{2011}, \emph{83} 035309.

\bibitem{Huang20sr}
Y.~Q. Huang, I.~A. Buyanova, W.~M. Chen,
\newblock \emph{Sci. Rep.} \textbf{{2020}}, \emph{{10}}, {1}.

\bibitem{McIver12natnano}
J.~W. McIver, D.~Hsieh, H.~Steinberg, P.~Jarillo-Herrero, N.~Gedik,
\newblock \emph{Nat. Nanotechnol.} \textbf{2012}, \emph{7} 96.

\bibitem{Duan14scirep}
J.~Duan, N.~Tang, X.~He, Y.~Yan, S.~Zhang, X.~Qin, X.~Wang, X.~Yang, F.~Xu,
  Y.~Chen, W.~Ge, B.~Shen,
\newblock \emph{Sci. Rep.} \textbf{{2014}}, \emph{{4}}.

\bibitem{Olbrich14prl}
P.~Olbrich, L.~E. Golub, T.~Herrmann, S.~N. Danilov, H.~Plank, V.~V. Bel'kov,
  G.~Mussler, C.~Weyrich, C.~M. Schneider, J.~Kampmeier, D.~Grützmacher,
  L.~Plucinski, M.~Eschbach, S.~D. Ganichev,
\newblock \emph{Phys. Rev. Lett.} \textbf{2014}, \emph{113} 096601.

\bibitem{Junck13prb}
A.~Junck, G.~Refael, F.~von Oppen,
\newblock \emph{Phys. Rev. B} \textbf{2013}, \emph{88} 075144.

\bibitem{Grinberg70jetp}
A.~A. Grinberg,
\newblock \emph{Sov. Phys. JETP} \textbf{{1970}}, \emph{{31}}, {3} {531}.

\bibitem{Plank16prb}
H.~Plank, L.~E. Golub, S.~Bauer, V.~V. Bel'kov, T.~Herrmann, P.~Olbrich,
  M.~Eschbach, L.~Plucinski, C.~M. Schneider, J.~Kampmeier, M.~Lanius,
  G.~Mussler, D.~Grützmacher, S.~D. Ganichev,
\newblock \emph{Phys. Rev. B} \textbf{{2016}}, \emph{{93}}, {12}.

\bibitem{Haight95ssr}
R.~Haight,
\newblock \emph{Surf. Sci. Rep.} \textbf{1995}, \emph{21} 277.

\bibitem{Fauster95}
T.~Fauster, W.~Steinmann,
\newblock In P.~Halevi, editor, \emph{Electromagnetic Waves: Recent
  Developments in Research}, volume~2, 347--411. North-Holland, Amsterdam,
  \textbf{1995}.

\bibitem{Bovens10}
U.~Bovensiepen, H.~Petek, M.~Wolf, editors,
\newblock \emph{Dynamics at Solid State Surfaces and Interfaces, {\rm Vol.~1:}
  Current Developments},
\newblock Wiley-VCH, Berlin, \textbf{2010}.

\bibitem{Berthold02prl}
W.~Berthold, U.~H{\"ofer}, P.~Feulner, E.~V. Chulkov, V.~M. Silkin, P.~M.
  Echenique,
\newblock \emph{Phys. Rev. Lett.} \textbf{2002}, \emph{88} 056805.

\bibitem{Rohleder05njp}
M.~Rohleder, K.~Duncker, W.~Berthold, J.~G{\"u}dde, U.~H{\"o}fer,
\newblock \emph{New J. Phys.} \textbf{2005}, \emph{7} 103.

\bibitem{Petek97pss}
H.~Petek, S.~Ogawa,
\newblock \emph{Prog. Surf. Sci.} \textbf{1997}, \emph{56} 239.

\bibitem{Gudde06apa}
J.~G{\"u}dde, M.~Rohleder, U.~H{\"o}fer,
\newblock \emph{Appl. Phys. A} \textbf{2006}, \emph{85} 345.

\bibitem{Reutzel20natcomm}
M.~Reutzel, A.~Li, Z.~Wang, H.~Petek,
\newblock \emph{Nat. Commun.} \textbf{2020}, \emph{11} 2230.

\bibitem{Wang12prl}
Y.~H. Wang, D.~Hsieh, E.~J. Sie, H.~Steinberg, D.~R. Gardner, Y.~S. Lee,
  P.~Jarillo-Herrero, N.~Gedik,
\newblock \emph{Phys. Rev. Lett.} \textbf{2012}, \emph{109} 127401.

\bibitem{Crepaldi12prb}
A.~Crepaldi, B.~Ressel, F.~Cilento, M.~Zacchigna, C.~Grazioli, H.~Berger,
  P.~Bugnon, K.~Kern, M.~Grioni, F.~Parmigiani,
\newblock \emph{Phys. Rev. B} \textbf{2012}, \emph{86} 205133.

\bibitem{Hajlaoui12nl}
M.~Hajlaoui, E.~Papalazarou, J.~Mauchain, G.~Lantz, N.~Moisan, D.~Boschetto,
  Z.~Jiang, I.~Miotkowski, Y.~P. Chen, A.~Taleb-Ibrahimi, L.~Perfetti,
  M.~Marsi,
\newblock \emph{Nano Lett.} \textbf{2012}, \emph{12} 3532.

\bibitem{Sobota12prl}
J.~A. Sobota, S.~Yang, J.~G. Analytis, Y.~L. Chen, I.~R. Fisher, P.~S.
  Kirchmann, Z.~X. Shen,
\newblock \emph{Phys. Rev. Lett.} \textbf{2012}, \emph{108} 117403.

\bibitem{Niesner12prb}
D.~Niesner, T.~Fauster, J.~I. Dadap, N.~Zaki, K.~R. Knox, P.~C. Yeh,
  R.~Bhandari, R.~M. Osgood, M.~Petrovic, M.~Kralj,
\newblock \emph{Phys. Rev. B} \textbf{2012}, \emph{85} 081402.

\bibitem{Crepaldi13prb}
A.~Crepaldi, F.~Cilento, B.~Ressel, C.~Cacho, J.~C. Johannsen, M.~Zacchigna,
  H.~Berger, P.~Bugnon, C.~Grazioli, I.~C.~E. Turcu, E.~Springate, K.~Kern,
  M.~Grioni, F.~Parmigiani,
\newblock \emph{Phys. Rev. B} \textbf{2013}, \emph{88} 121404.

\bibitem{Hajlaoui13epj}
M.~Hajlaoui, E.~Papalazarou, J.~Mauchain, Z.~Jiang, I.~Miotkowski, Y.~P. Chen,
  A.~Taleb-Ibrahimi, L.~Perfetti, M.~Marsi,
\newblock \emph{Eur. Phys. J.-Spec. Top.} \textbf{2013}, \emph{222} 1271.

\bibitem{Sobota13prl}
J.~A. Sobota, S.~L. Yang, A.~F. Kemper, J.~J. Lee, F.~T. Schmitt, W.~Li, R.~G.
  Moore, J.~G. Analytis, I.~R. Fisher, P.~S. Kirchmann, T.~P. Devereaux, Z.~X.
  Shen,
\newblock \emph{Phys. Rev. Lett.} \textbf{2013}, \emph{111} 136802.

\bibitem{Niesner14prb}
D.~Niesner, S.~Otto, V.~Hermann, T.~Fauster, T.~V. Menshchikova, S.~V. Eremeev,
  Z.~S. Aliev, I.~R. Amiraslanov, M.~B. Babanly, P.~M. Echenique, E.~V.
  Chulkov,
\newblock \emph{Phys. Rev. B} \textbf{2014}, \emph{89} 081404.
Erratum \newblock \emph{Phys. Rev. B} \textbf{2015}, \emph{91} 039903.

\bibitem{Hajlaoui14natcomm}
M.~Hajlaoui, E.~Papalazarou, J.~Mauchain, L.~Perfetti, A.~Taleb-Ibrahimi,
  F.~Navarin, M.~Monteverde, P.~Auban-Senzier, C.~R. Pasquier, N.~Moisan,
  D.~Boschetto, M.~Neupane, M.~Z. Hasan, T.~Durakiewicz, Z.~Jiang, Y.~Xu,
  I.~Miotkowski, Y.~P. Chen, S.~Jia, H.~W. Ji, R.~J. Cava, M.~Marsi,
\newblock \emph{Nat. Commun.} \textbf{2014}, \emph{5} 3003.

\bibitem{Reimann14prb}
J.~Reimann, J.~G{\"u}dde, K.~Kuroda, E.~V. Chulkov, U.~H{\"o}fer,
\newblock \emph{Phys. Rev. B} \textbf{2014}, \emph{90} 081106.
\newblock Erratum: \emph{Phys. Rev. B} \textbf{2015}, \emph{91} 039903.

\bibitem{Neupane15prl}
M.~Neupane, S.~Y. Xu, Y.~Ishida, S.~Jia, B.~M. Fregoso, C.~Liu, I.~Belopolski,
  G.~Bian, N.~Alidoust, T.~Durakiewicz, V.~Galitski, S.~Shin, R.~J. Cava, M.~Z.
  Hasan,
\newblock \emph{Phys. Rev. Lett.} \textbf{2015}, \emph{115} 116801.

\bibitem{Zhu15sr2}
S.~Zhu, Y.~Ishida, K.~Kuroda, K.~Sumida, M.~Ye, J.~Wang, H.~Pan, M.~Taniguchi,
  S.~Qiao, S.~Shin, A.~Kimura,
\newblock \emph{Sci. Rep.} \textbf{2015}, \emph{5} 13213.

\bibitem{Sanchez16prb}
J.~S{\'a}nchez-Barriga, E.~Golias, A.~Varykhalov, J.~Braun, L.~V. Yashina,
  R.~Schumann, J.~Minar, H.~Ebert, O.~Kornilov, O.~Rader,
\newblock \emph{Phys. Rev. B} \textbf{2016}, \emph{93} 155426.

\bibitem{Jozwiak16natcomm}
C.~Jozwiak, J.~A. Sobota, K.~Gotlieb, A.~F. Kemper, C.~R. Rotundu, R.~J.
  Birgeneau, Z.~Hussain, D.-H. Lee, Z.-X. Shen, A.~Lanzara,
\newblock \emph{Nature Commun.} \textbf{{2016}}, \emph{{7}}.

\bibitem{Sumida17scirep}
K.~Sumida, Y.~Ishida, S.~Zhu, M.~Ye, A.~Pertsova, C.~Triola, K.~A. Kokh, O.~E.
  Tereshchenko, A.~V. Balatsky, S.~Shin, A.~Kimura,
\newblock \emph{Sci. Rep.} \textbf{2017}, \emph{7}, 1 14080.

\bibitem{Bugini17jpcm}
D.~Bugini, F.~Boschini, H.~Hedayat, H.~Yi, C.~Chen, X.~Zhou, C.~Manzoni,
  C.~Dallera, G.~Cerullo, E.~Carpene,
\newblock \emph{{J. Phys. Condens. Matter}} \textbf{{2017}}, \emph{{29}}, {30}.

\bibitem{Sanchez17prb}
J.~S{\'a}nchez-Barriga, M.~Battiato, M.~Krivenkov, E.~Golias, A.~Varykhalov,
  A.~Romualdi, L.~V. Yashina, J.~Min\'ar, O.~Kornilov, H.~Ebert, K.~Held,
  J.~Braun,
\newblock \emph{Phys. Rev. B} \textbf{2017}, \emph{95} 125405.

\bibitem{Ketterl18prb}
A.~S. Ketterl, S.~Otto, M.~Bastian, B.~Andres, C.~Gahl, J.~Minar, H.~Ebert,
  J.~Braun, O.~E. Tereshchenko, K.~A. Kokh, T.~Fauster, M.~Weinelt,
\newblock \emph{Phys. Rev. B} \textbf{2018}, \emph{98} 155406.

\bibitem{Soifer19prl}
H.~Soifer, A.~Gauthier, A.~F. Kemper, C.~R. Rotundu, S.-L. Yang, H.~Xiong,
  D.~Lu, M.~Hashimoto, P.~S. Kirchmann, J.~A. Sobota, Z.-X. Shen,
\newblock \emph{Phys. Rev. Lett.} \textbf{2019}, \emph{122} 167401.

\bibitem{Horak88pss}
J.~Hor{\'a}k, Z.~Star{\'y}, J.~Klikorka,
\newblock \emph{phys. stat. sol. (b)} \textbf{1988}, \emph{147} 501.

\bibitem{Pauly12prb}
C.~Pauly, G.~Bihlmayer, M.~Liebmann, M.~Grob, A.~Georgi, D.~Subramaniam, M.~R.
  Scholz, J.~S{\'a}nchez-Barriga, A.~Varykhalov, S.~Blugel, O.~Rader,
  M.~Morgenstern,
\newblock \emph{Phys. Rev. B} \textbf{2012}, \emph{86} 235106.

\bibitem{Seibel12prb}
C.~Seibel, H.~Maass, M.~Ohtak, S.~Fiedler, C.~Junger, C.~H. Min, H.~Bentmann,
  K.~Sakamoto, F.~Reinert,
\newblock \emph{Phys. Rev. B} \textbf{2012}, \emph{86} 161105.

\bibitem{Jiang12prl}
Y.~Jiang, Y.~Wang, M.~Chen, Z.~Li, C.~Song, K.~He, L.~Wang, X.~Chen, X.~Ma,
  Q.-K. Xue,
\newblock \emph{Phys. Rev. Lett.} \textbf{2012}, \emph{108} 016401.

\bibitem{Menshc11jetpl}
T.~V. Menshchikova, S.~V. Eremeev, E.~V. Chulkov,
\newblock \emph{JETP Lett.} \textbf{2011}, \emph{94} 106.

\bibitem{Lin11njp}
H.~Lin, T.~Das, L.~A. Wray, S.~Y. Xu, M.~Z. Hasan, A.~Bansil,
\newblock \emph{New J. Phys.} \textbf{2011}, \emph{13} 095005.

\bibitem{Glazov14}
M.~M. Glazov, S.~D. Ganichev,
\newblock \emph{Phys. Rep.} \textbf{2014}, \emph{535} 101.

\bibitem{Kastl15natcomm}
C.~Kastl, C.~Karnetzky, H.~Karl, A.~W. Holleitner,
\newblock \emph{Nat. Commun.} \textbf{2015}, \emph{6} 6617.

\bibitem{Pan17natcomm}
Y.~Pan, Q.-Z. Wang, A.~L. Yeats, T.~Pillsbury, T.~C. Flanagan, A.~Richardella,
  H.~Zhang, D.~D. Awschalom, C.-X. Liu, N.~Samarth,
\newblock \emph{Nature Commun.} \textbf{{2017}}, \emph{{8}}.

\bibitem{Kuroda16prl}
K.~Kuroda, J.~Reimann, J.~G{\"u}dde, U.~H{\"o}fer,
\newblock \emph{Phys. Rev. Lett.} \textbf{2016}, \emph{116} 076801.

\bibitem{Moser17jelsp}
S.~Moser,
\newblock \emph{J. Electron Spectrosc.} \textbf{2017}, \emph{214} 29.

\bibitem{Glinka15prb}
Y.~D. Glinka, S.~Babakiray, T.~A. Johnson, M.~B. Holcomb, D.~Lederman,
\newblock \emph{Phys. Rev. B} \textbf{2015}, \emph{91} 195307.

\bibitem{Kuroda17spie}
K.~Kuroda, J.~Reimann, J.~G{\"u}dde, U.~H{\"o}fer,
\newblock \emph{Proc. SPIE 10102, Ultrafast Phenomena and
  Nanophotonics XXI} \textbf{2017} \emph{101020} 71.

\bibitem{Fauster07pss}
T.~Fauster, M.~Weinelt, U.~H{\"o}fer,
\newblock \emph{Prog. Surf. Sci.} \textbf{2007}, \emph{82} 224.

\bibitem{Marks11prb2}
M.~Marks, C.~H. Schwalb, K.~Schubert, J.~G{\"u}dde, U.~H{\"o}fer,
\newblock \emph{Phys. Rev. B} \textbf{2011}, \emph{84} 245402.

\bibitem{Kuroda17prb}
K.~Kuroda, J.~Reimann, K.~A. Kokh, O.~E. Tereshchenko, A.~Kimura, J.~G{\"u}dde,
  U.~H{\"o}fer,
\newblock \emph{Phys. Rev. B} \textbf{2017}, \emph{95} 081103(R).

\bibitem{Reimann18nat}
J.~Reimann, S.~Schlauderer, C.~P. Schmid, F.~Langer, S.~Baierl, K.~A. Kokh,
  O.~E. Tereshchenko, A.~Kimura, C.~Lange, G{\"u}dde, U.~H{\"o}fer, R.~Huber,
\newblock \emph{Nature} \textbf{2018}, \emph{562} 396.

\bibitem{Tokura19natrev}
Y.~Tokura, K.~Yasuda, A.~Tsukazaki,
\newblock \emph{Nat. Rev. Phys.} \textbf{{2019}}, \emph{{1}} 126.

\end{thebibliography}

\end{document}